\documentclass[draftcls,10pt,onecolumn]{IEEEtran}
\usepackage{cite}
\usepackage[T1]{fontenc}
\usepackage{graphicx}

\usepackage{amsmath,amssymb,amsfonts,amsthm}
\usepackage{paralist}
\usepackage{subfigure}
\usepackage{setspace}
\usepackage{mathpazo}
\usepackage{microtype}

\usepackage{algorithm}
\usepackage[hidelinks]{hyperref}
\hypersetup{pdftitle={Recovery of Signals with Low Density},pdfauthor={Christoph Studer}}
\usepackage{algpseudocode}

\usepackage{amssymb}
\usepackage{amsfonts}
\usepackage{mathrsfs}
\usepackage{xspace}
\usepackage{bm}
\usepackage{upgreek}

\newcommand{\safemath}[2]{\newcommand{#1}{\ensuremath{#2}\xspace}}

\safemath{\bma}{\mathbf{a}}
\safemath{\bmb}{\mathbf{b}}
\safemath{\bmc}{\mathbf{c}}
\safemath{\bmd}{\mathbf{d}}
\safemath{\bme}{\mathbf{e}}
\safemath{\bmf}{\mathbf{f}}
\safemath{\bmg}{\mathbf{g}}
\safemath{\bmh}{\mathbf{h}}
\safemath{\bmi}{\mathbf{i}}
\safemath{\bmj}{\mathbf{j}}
\safemath{\bmk}{\mathbf{k}}
\safemath{\bml}{\mathbf{l}}
\safemath{\bmm}{\mathbf{m}}
\safemath{\bmn}{\mathbf{n}}
\safemath{\bmo}{\mathbf{o}}
\safemath{\bmp}{\mathbf{p}}
\safemath{\bmq}{\mathbf{q}}
\safemath{\bmr}{\mathbf{r}}
\safemath{\bms}{\mathbf{s}}
\safemath{\bmt}{\mathbf{t}}
\safemath{\bmu}{\mathbf{u}}
\safemath{\bmv}{\mathbf{v}}
\safemath{\bmw}{\mathbf{w}}
\safemath{\bmx}{\mathbf{x}}
\safemath{\bmy}{\mathbf{y}}
\safemath{\bmz}{\mathbf{z}}
\safemath{\bmzero}{\mathbf{0}}
\safemath{\bmone}{\mathbf{1}}

\bmdefine{\biad}{a}
\bmdefine{\bibd}{b}
\bmdefine{\bicd}{c}
\bmdefine{\bidd}{d}
\bmdefine{\bied}{e}
\bmdefine{\bifd}{f}
\bmdefine{\bigd}{g}
\bmdefine{\bihd}{h}
\bmdefine{\biid}{i}
\bmdefine{\bijd}{j}
\bmdefine{\bikd}{k}
\bmdefine{\bild}{l}
\bmdefine{\bimd}{m}
\bmdefine{\bind}{n}
\bmdefine{\biod}{o}
\bmdefine{\bipd}{p}
\bmdefine{\biqd}{q}
\bmdefine{\bird}{r}
\bmdefine{\bisd}{s}
\bmdefine{\bitd}{t}
\bmdefine{\biud}{u}
\bmdefine{\bivd}{v}
\bmdefine{\biwd}{w}
\bmdefine{\bixd}{x}
\bmdefine{\biyd}{y}
\bmdefine{\bizd}{z}

\bmdefine{\bixid}{\xi}
\bmdefine{\bilambdad}{\lambda}
\bmdefine{\bimud}{\mu}
\bmdefine{\bithetad}{\theta}
\bmdefine{\biphid}{\phi}
\bmdefine{\bideltad}{\delta}

\safemath{\bmia}{\biad}
\safemath{\bmib}{\bibd}
\safemath{\bmic}{\bicd}
\safemath{\bmid}{\bidd}
\safemath{\bmie}{\bied}
\safemath{\bmif}{\bifd}
\safemath{\bmig}{\bigd}
\safemath{\bmih}{\bihd}
\safemath{\bmii}{\biid}
\safemath{\bmij}{\bijd}
\safemath{\bmik}{\bikd}
\safemath{\bmil}{\bild}
\safemath{\bmim}{\bimd}
\safemath{\bmin}{\bind}
\safemath{\bmio}{\biod}
\safemath{\bmip}{\bipd}
\safemath{\bmiq}{\biqd}
\safemath{\bmir}{\bird}
\safemath{\bmis}{\bisd}
\safemath{\bmit}{\bitd}
\safemath{\bmiu}{\biud}
\safemath{\bmiv}{\bivd}
\safemath{\bmiw}{\biwd}
\safemath{\bmix}{\bixd}
\safemath{\bmiy}{\biyd}
\safemath{\bmiz}{\bizd}

\safemath{\bmxi}{\bixid}
\safemath{\bmlambda}{\bilambdad}
\safemath{\bmmu}{\bimud}
\safemath{\bmtheta}{\bithetad}
\safemath{\bmphi}{\biphid}
\safemath{\bmdelta}{\bideltad}

\safemath{\bA}{\mathbf{A}}
\safemath{\bB}{\mathbf{B}}
\safemath{\bC}{\mathbf{C}}
\safemath{\bD}{\mathbf{D}}
\safemath{\bE}{\mathbf{E}}
\safemath{\bF}{\mathbf{F}}
\safemath{\bG}{\mathbf{G}}
\safemath{\bH}{\mathbf{H}}
\safemath{\bI}{\mathbf{I}}
\safemath{\bJ}{\mathbf{J}}
\safemath{\bK}{\mathbf{K}}
\safemath{\bL}{\mathbf{L}}
\safemath{\bM}{\mathbf{M}}
\safemath{\bN}{\mathbf{N}}
\safemath{\bO}{\mathbf{O}}
\safemath{\bP}{\mathbf{P}}
\safemath{\bQ}{\mathbf{Q}}
\safemath{\bR}{\mathbf{R}}
\safemath{\bS}{\mathbf{S}}
\safemath{\bT}{\mathbf{T}}
\safemath{\bU}{\mathbf{U}}
\safemath{\bV}{\mathbf{V}}
\safemath{\bW}{\mathbf{W}}
\safemath{\bX}{\mathbf{X}}
\safemath{\bY}{\mathbf{Y}}
\safemath{\bZ}{\mathbf{Z}}

\safemath{\bZero}{\mathbf{0}}
\safemath{\bOne}{\mathbf{1}}
\safemath{\bDelta}{\mathbf{\Delta}}
\safemath{\bLambda}{\mathbf{\UpLambda}}
\safemath{\bPhi}{\mathbf{\Upphi}}
\safemath{\bSigma}{\mathbf{\Upsigma}}
\safemath{\bOmega}{\mathbf{\Upomega}}
\safemath{\bTheta}{\mathbf{\Uptheta}}

\bmdefine{\biAd}{A}
\bmdefine{\biBd}{B}
\bmdefine{\biCd}{C}
\bmdefine{\biDd}{D}
\bmdefine{\biEd}{E}
\bmdefine{\biFd}{F}
\bmdefine{\biGd}{G}
\bmdefine{\biHd}{H}
\bmdefine{\biId}{I}
\bmdefine{\biJd}{J}
\bmdefine{\biKd}{K}
\bmdefine{\biLd}{L}
\bmdefine{\biMd}{M}
\bmdefine{\biOd}{N}
\bmdefine{\biPd}{O}
\bmdefine{\biQd}{P}
\bmdefine{\biRd}{R}
\bmdefine{\biSd}{S}
\bmdefine{\biTd}{T}
\bmdefine{\biUd}{U}
\bmdefine{\biVd}{V}
\bmdefine{\biWd}{W}
\bmdefine{\biXd}{X}
\bmdefine{\biYd}{Y}
\bmdefine{\biZd}{Z}

\bmdefine{\biDelta}{\Delta}
\bmdefine{\biLambda}{\Lambda}
\bmdefine{\biPhi}{\Phi}
\bmdefine{\biSigma}{\Sigma}
\bmdefine{\biOmega}{\Omega}
\bmdefine{\biTheta}{\Theta}

\safemath{\bimA}{\biAd}
\safemath{\bimB}{\biBd}
\safemath{\bimC}{\biCd}
\safemath{\bimD}{\biDd}
\safemath{\bimE}{\biEd}
\safemath{\bimF}{\biFd}
\safemath{\bimG}{\biGd}
\safemath{\bimH}{\biHd}
\safemath{\bimI}{\biId}
\safemath{\bimJ}{\biJd}
\safemath{\bimK}{\biKd}
\safemath{\bimL}{\biLd}
\safemath{\bimM}{\biMd}
\safemath{\bimN}{\biNd}
\safemath{\bimO}{\biOd}
\safemath{\bimP}{\biPd}
\safemath{\bimQ}{\biQd}
\safemath{\bimR}{\biRd}
\safemath{\bimS}{\biSd}
\safemath{\bimT}{\biTd}
\safemath{\bimU}{\biUd}
\safemath{\bimV}{\biVd}
\safemath{\bimW}{\biWd}
\safemath{\bimX}{\biXd}
\safemath{\bimY}{\biYd}
\safemath{\bimZ}{\biZd}

\safemath{\bimDelta}{\biDelta}
\safemath{\bimLambda}{\biLambda}
\safemath{\bimPhi}{\biPhi}
\safemath{\bimSigma}{\biSigma}
\safemath{\bimOmega}{\biOmega}
\safemath{\bimTheta}{\biTheta}

\safemath{\setA}{\mathcal{A}}
\safemath{\setB}{\mathcal{B}}
\safemath{\setC}{\mathcal{C}}
\safemath{\setD}{\mathcal{D}}
\safemath{\setE}{\mathcal{E}}
\safemath{\setF}{\mathcal{F}}
\safemath{\setG}{\mathcal{G}}
\safemath{\setH}{\mathcal{H}}
\safemath{\setI}{\mathcal{I}}
\safemath{\setJ}{\mathcal{J}}
\safemath{\setK}{\mathcal{K}}
\safemath{\setL}{\mathcal{L}}
\safemath{\setM}{\mathcal{M}}
\safemath{\setN}{\mathcal{N}}
\safemath{\setO}{\mathcal{O}}
\safemath{\setP}{\mathcal{P}}
\safemath{\setQ}{\mathcal{Q}}
\safemath{\setR}{\mathcal{R}}
\safemath{\setS}{\mathcal{S}}
\safemath{\setT}{\mathcal{T}}
\safemath{\setU}{\mathcal{U}}
\safemath{\setV}{\mathcal{V}}
\safemath{\setW}{\mathcal{W}}
\safemath{\setX}{\mathcal{X}}
\safemath{\setY}{\mathcal{Y}}
\safemath{\setZ}{\mathcal{Z}}
\safemath{\emptySet}{\varnothing}

\safemath{\colA}{\mathscr{A}}
\safemath{\colB}{\mathscr{B}}
\safemath{\colC}{\mathscr{C}}
\safemath{\colD}{\mathscr{D}}
\safemath{\colE}{\mathscr{E}}
\safemath{\colF}{\mathscr{F}}
\safemath{\colG}{\mathscr{G}}
\safemath{\colH}{\mathscr{H}}
\safemath{\colI}{\mathscr{I}}
\safemath{\colJ}{\mathscr{J}}
\safemath{\colK}{\mathscr{K}}
\safemath{\colL}{\mathscr{L}}
\safemath{\colM}{\mathscr{M}}
\safemath{\colN}{\mathscr{N}}
\safemath{\colO}{\mathscr{O}}
\safemath{\colP}{\mathscr{P}}
\safemath{\colQ}{\mathscr{Q}}
\safemath{\colR}{\mathscr{R}}
\safemath{\colS}{\mathscr{S}}
\safemath{\colT}{\mathscr{T}}
\safemath{\colU}{\mathscr{U}}
\safemath{\colV}{\mathscr{V}}
\safemath{\colW}{\mathscr{W}}
\safemath{\colX}{\mathscr{X}}
\safemath{\colY}{\mathscr{Y}}
\safemath{\colZ}{\mathscr{Z}}

\safemath{\opA}{\mathbb{A}}
\safemath{\opB}{\mathbb{B}}
\safemath{\opC}{\mathbb{C}}
\safemath{\opD}{\mathbb{D}}
\safemath{\opE}{\mathbb{E}}
\safemath{\opF}{\mathbb{F}}
\safemath{\opG}{\mathbb{G}}
\safemath{\opH}{\mathbb{H}}
\safemath{\opI}{\mathbb{I}}
\safemath{\opJ}{\mathbb{J}}
\safemath{\opK}{\mathbb{K}}
\safemath{\opL}{\mathbb{L}}
\safemath{\opM}{\mathbb{M}}
\safemath{\opN}{\mathbb{N}}
\safemath{\opO}{\mathbb{O}}
\safemath{\opP}{\mathbb{P}}
\safemath{\opQ}{\mathbb{Q}}
\safemath{\opR}{\mathbb{R}}
\safemath{\opS}{\mathbb{S}}
\safemath{\opT}{\mathbb{T}}
\safemath{\opU}{\mathbb{U}}
\safemath{\opV}{\mathbb{V}}
\safemath{\opW}{\mathbb{W}}
\safemath{\opX}{\mathbb{X}}
\safemath{\opY}{\mathbb{Y}}
\safemath{\opZ}{\mathbb{Z}}
\safemath{\opZero}{\mathbb{O}}
\safemath{\identityop}{\opI}

\safemath{\veca}{\bma}
\safemath{\vecb}{\bmb}
\safemath{\vecc}{\bmc}
\safemath{\vecd}{\bmd}
\safemath{\vece}{\bme}
\safemath{\vecf}{\bmf}
\safemath{\vecg}{\bmg}
\safemath{\vech}{\bmh}
\safemath{\veci}{\bmi}
\safemath{\vecj}{\bmj}
\safemath{\veck}{\bmk}
\safemath{\vecl}{\bml}
\safemath{\vecm}{\bmm}
\safemath{\vecn}{\bmn}
\safemath{\veco}{\bmo}
\safemath{\vecp}{\bmp}
\safemath{\vecq}{\bmq}
\safemath{\vecr}{\bmr}
\safemath{\vecs}{\bms}
\safemath{\vect}{\bmt}
\safemath{\vecu}{\bmu}
\safemath{\vecv}{\bmv}
\safemath{\vecw}{\bmw}
\safemath{\vecx}{\bmx}
\safemath{\vecy}{\bmy}
\safemath{\vecz}{\bmz}

\safemath{\veczero}{\bmzero}
\safemath{\vecone}{\bmone}
\safemath{\vecxi}{\bmxi}
\safemath{\veclambda}{\bmlambda}
\safemath{\vecmu}{\bmmu}
\safemath{\vectheta}{\bmtheta}
\safemath{\vecphi}{\bmphi}
\safemath{\vecdelta}{\bmdelta}

\safemath{\matA}{\bA}
\safemath{\matB}{\bB}
\safemath{\matC}{\bC}
\safemath{\matD}{\bD}
\safemath{\matE}{\bE}
\safemath{\matF}{\bF}
\safemath{\matG}{\bG}
\safemath{\matH}{\bH}
\safemath{\matI}{\bI}
\safemath{\matJ}{\bJ}
\safemath{\matK}{\bK}
\safemath{\matL}{\bL}
\safemath{\matM}{\bM}
\safemath{\matN}{\bN}
\safemath{\matO}{\bO}
\safemath{\matP}{\bP}
\safemath{\matQ}{\bQ}
\safemath{\matR}{\bR}
\safemath{\matS}{\bS}
\safemath{\matT}{\bT}
\safemath{\matU}{\bU}
\safemath{\matV}{\bV}
\safemath{\matW}{\bW}
\safemath{\matX}{\bX}
\safemath{\matY}{\bY}
\safemath{\matZ}{\bZ}
\safemath{\matzero}{\bmzero}

\safemath{\matDelta}{\bDelta}
\safemath{\matLambda}{\bLambda}
\safemath{\matPhi}{\bPhi}
\safemath{\matSigma}{\bSigma}
\safemath{\matOmega}{\bOmega}
\safemath{\matTheta}{\bTheta}

\safemath{\matidentity}{\matI}
\safemath{\matone}{\matO}

\safemath{\rnda}{A}
\safemath{\rndb}{B}
\safemath{\rndc}{C}
\safemath{\rndd}{D}
\safemath{\rnde}{E}
\safemath{\rndf}{F}
\safemath{\rndg}{G}
\safemath{\rndh}{H}
\safemath{\rndi}{I}
\safemath{\rndj}{J}
\safemath{\rndk}{K}
\safemath{\rndl}{L}
\safemath{\rndm}{M}
\safemath{\rndn}{N}
\safemath{\rndo}{O}
\safemath{\rndp}{P}
\safemath{\rndq}{Q}
\safemath{\rndr}{R}
\safemath{\rnds}{S}
\safemath{\rndt}{T}
\safemath{\rndu}{U}
\safemath{\rndv}{V}
\safemath{\rndw}{W}
\safemath{\rndx}{X}
\safemath{\rndy}{Y}
\safemath{\rndz}{Z}

\safemath{\rveca}{\bimA}
\safemath{\rvecb}{\bimB}
\safemath{\rvecc}{\bimC}
\safemath{\rvecd}{\bimD}
\safemath{\rvece}{\bimE}
\safemath{\rvecf}{\bimF}
\safemath{\rvecg}{\bimG}
\safemath{\rvech}{\bimH}
\safemath{\rveci}{\bimI}
\safemath{\rvecj}{\bimJ}
\safemath{\rveck}{\bimK}
\safemath{\rvecl}{\bimL}
\safemath{\rvecm}{\bimM}
\safemath{\rvecn}{\bimN}
\safemath{\rveco}{\bomO}
\safemath{\rvecp}{\bimP}
\safemath{\rvecq}{\bimQ}
\safemath{\rvecr}{\bimR}
\safemath{\rvecs}{\bimS}
\safemath{\rvect}{\bimT}
\safemath{\rvecu}{\bimU}
\safemath{\rvecv}{\bimV}
\safemath{\rvecw}{\bimW}
\safemath{\rvecx}{\bimX}
\safemath{\rvecy}{\bimY}
\safemath{\rvecz}{\bimZ}

\safemath{\rvecxi}{\bmxi}
\safemath{\rveclambda}{\bmlambda}
\safemath{\rvecmu}{\bmmu}
\safemath{\rvectheta}{\bmtheta}
\safemath{\rvecphi}{\bmphi}

\safemath{\rmatA}{\bimA}
\safemath{\rmatB}{\bimB}
\safemath{\rmatC}{\bimC}
\safemath{\rmatD}{\bimD}
\safemath{\rmatE}{\bimE}
\safemath{\rmatF}{\bimF}
\safemath{\rmatG}{\bimG}
\safemath{\rmatH}{\bimH}
\safemath{\rmatI}{\bimI}
\safemath{\rmatJ}{\bimJ}
\safemath{\rmatK}{\bimK}
\safemath{\rmatL}{\bimL}
\safemath{\rmatM}{\bimM}
\safemath{\rmatN}{\bimN}
\safemath{\rmatO}{\bimO}
\safemath{\rmatP}{\bimP}
\safemath{\rmatQ}{\bimQ}
\safemath{\rmatR}{\bimR}
\safemath{\rmatS}{\bimS}
\safemath{\rmatT}{\bimT}
\safemath{\rmatU}{\bimU}
\safemath{\rmatV}{\bimV}
\safemath{\rmatW}{\bimW}
\safemath{\rmatX}{\bimX}
\safemath{\rmatY}{\bimY}
\safemath{\rmatZ}{\bimZ}

\safemath{\rmatDelta}{\bimDelta}
\safemath{\rmatLambda}{\bimLambda}
\safemath{\rmatPhi}{\bimPhi}
\safemath{\rmatSigma}{\bimSigma}
\safemath{\rmatOmega}{\bimOmega}
\safemath{\rmatTheta}{\bimTheta}

\usepackage{amssymb}
\usepackage{amsfonts}
\usepackage{mathrsfs}
\usepackage{xspace}
\usepackage{bm}
\usepackage{fancyref}
\usepackage{textcomp}

\usepackage{multirow}
\usepackage{stmaryrd}

\newenvironment{textbmatrix}{	\setlength{\arraycolsep}{2.5pt}%
								\big[\begin{matrix}}{\end{matrix}\big]%
								\raisebox{0.08ex}{\vphantom{M}}}

\def\be{\begin{equation}}
\def\ee{\end{equation}}
\def\een{\nonumber \end{equation}}
\def\mat{\begin{bmatrix}}
\def\emat{\end{bmatrix}}
\def\btm{\begin{textbmatrix}}
\def\etm{\end{textbmatrix}}

\def\ba#1\ea{\begin{align}#1\end{align}}
\def\bas#1\eas{\begin{align*}#1\end{align*}}
\def\bs#1\es{\begin{split}#1\end{split}} 
\def\bg#1\eg{\begin{gather}#1\end{gather}}
\def\bml#1\eml{\begin{multline}#1\end{multline}}
\def\bi#1\ei{\begin{itemize}#1\end{itemize}}

\newcommand{\lefto}{\mathopen{}\left}

\newcommand{\abs}[1]{\lefto\lvert#1\right\rvert}		

\newcommand{\vecnorm}[1]{\lefto\lVert#1\right\rVert}		
\newcommand{\herm}{\ensuremath{\textnormal{H}}} 	

\safemath{\dirac}{\delta}					
\safemath{\krond}{\dirac}					

\safemath{\upto}{\uparrow}
\safemath{\downto}{\downarrow}
\safemath{\iu}{j}							
\safemath{\ev}{\lambda}						
\safemath{\hilseqspace}{l^{2}}				
\newcommand{\banachfunspace}[1]{\setL^{#1}}	
\safemath{\hilfunspace}{\banachfunspace{2}}	

\safemath{\SNR}{\textsf{SNR}} 				
\safemath{\PAR}{\textsf{PAR}} 				
\safemath{\No}{N_0}							
\safemath{\Es}{E_s}							
\safemath{\Eb}{E_b}							
\safemath{\EbNo}{\frac{\Eb}{\No}}
\safemath{\EsNo}{\frac{\Es}{\No}}

\DeclareMathOperator{\CHop}{\ensuremath{\opH}} 
\safemath{\tvir}{\rndh_{\CHop}}				
\safemath{\tvtf}{\rndl_{\CHop}}				
\safemath{\spf}{\rnds_{\CHop}}				
\safemath{\bff}{H_{\CHop}}					

\safemath{\ircf}{r_{h}}						
\safemath{\tftvcf}{r_{s}}					
\safemath{\tfcf}{r_{l}}						
\safemath{\bfcf}{r_{H}}						

\safemath{\tcorr}{c_h}						
\safemath{\scf}{c_{s}}						
\safemath{\tfcorr}{c_{l}}					
\safemath{\fcorr}{c_{H}}						

\safemath{\mi}{I}							
\safemath{\capacity}{C}						

\safemath{\normal}{\mathcal{N}}			
\safemath{\jpg}{\mathcal{CN}}			
\safemath{\mchain}{\leftrightarrow}		

\safemath{\dB}{\,\mathrm{dB}}
\safemath{\dBm}{\,\mathrm{dBm}}
\safemath{\Hz}{\,\mathrm{Hz}}
\safemath{\kHz}{\,\mathrm{kHz}}
\safemath{\MHz}{\,\mathrm{MHz}}
\safemath{\GHz}{\,\mathrm{GHz}}
\safemath{\s}{\,\mathrm{s}}
\safemath{\ms}{\,\mathrm{ms}}
\safemath{\mus}{\,\mathrm{\text{\textmu}s}}
\safemath{\ns}{\,\mathrm{ns}}
\safemath{\ps}{\,\mathrm{ps}}
\safemath{\meter}{\,\mathrm{m}}
\safemath{\mm}{\,\mathrm{mm}}
\safemath{\cm}{\,\mathrm{cm}}
\safemath{\m}{\,\mathrm{m}}
\safemath{\W}{\,\mathrm{W}}
\safemath{\mW}{\, \mathrm{mW}}
\safemath{\J}{\,\mathrm{J}}
\safemath{\K}{\,\mathrm{K}}
\safemath{\bit}{\,\mathrm{bit}}
\safemath{\nat}{\,\mathrm{nat}}

\safemath{\define}{\triangleq}			

\safemath{\equivalent}{\sim}
\safemath{\distas}{\sim}					
\safemath{\sdiff}{\Delta}				

\safemath{\reals}{\mathbb{R}}
\safemath{\positivereals}{\reals_{+}}
\safemath{\integers}{\mathbb{Z}}
\safemath{\posint}{\integers_{+}}
\safemath{\naturals}{\mathbb{N}}
\safemath{\posnaturals}{\naturals_{+}}
\safemath{\complexset}{\mathbb{C}}
\safemath{\rationals}{\mathbb{Q}}

\newcommand*{\fancyrefapplabelprefix}{app}		
\newcommand*{\fancyrefthmlabelprefix}{thm}		
\newcommand*{\fancyreflemlabelprefix}{lem}		
\newcommand*{\fancyrefcorlabelprefix}{cor}		
\newcommand*{\fancyrefdeflabelprefix}{def}		
\newcommand*{\fancyrefproplabelprefix}{prop}		
\newcommand*{\fancyrefobslabelprefix}{obs}		
\newcommand*{\fancyrefexmpllabelprefix}{exmpl}
\newcommand*{\fancyrefalglabelprefix}{alg}		
\newcommand*{\fancyrefremlabelprefix}{rem}		

\frefformat{vario}{\fancyrefseclabelprefix}{Section~#1}
\frefformat{vario}{\fancyrefthmlabelprefix}{Theorem~#1}
\frefformat{vario}{\fancyreflemlabelprefix}{Lemma~#1}
\frefformat{vario}{\fancyrefcorlabelprefix}{Corollary~#1}
\frefformat{vario}{\fancyrefdeflabelprefix}{Definition~#1}
\frefformat{vario}{\fancyrefobslabelprefix}{Observation~#1}
\frefformat{vario}{\fancyreffiglabelprefix}{Fig.~#1}
\frefformat{vario}{\fancyrefapplabelprefix}{Appendix~#1} 
\frefformat{vario}{\fancyrefeqlabelprefix}{(#1)}
\frefformat{vario}{\fancyrefproplabelprefix}{Proposition~#1}
\frefformat{vario}{\fancyrefexmpllabelprefix}{Example~#1}
\frefformat{vario}{\fancyrefalglabelprefix}{Algorithm~#1}
\frefformat{vario}{\fancyrefremlabelprefix}{Remark~#1}

\newtheorem{thm}{Theorem}
\newtheorem{cor}[thm]{Corollary}   

\newtheorem{defi}{Definition}
\newtheorem{lem}[thm]{Lemma} 

\safemath{\dictab}{[\,\dicta\,\,\dictb\,]}

\safemath{\ysig}{\bmy}
\safemath{\ysighat}{\hat{\ysig}}
\safemath{\ysigdim}{M}
\safemath{\xsig}{\bmx}
\safemath{\xsigdim}{N}
\safemath{\nx}{n_x}
\safemath{\zsig}{\bmz}
\safemath{\zsigdim}{\ysigdim}
\safemath{\rsig}{\bmr}
\safemath{\Adict}{\bA}
\safemath{\Adicttilde}{\widetilde{\Adict}}
\safemath{\Adictdim}{\outputdim\times\xsigdim}
\safemath{\avec}{\bma}
\safemath{\avectilde}{\tilde{\avec}}
\safemath{\Bdict}{\bB}
\safemath{\Bdicttilde}{\widetilde{\Bdict}}
\safemath{\Cdict}{\bC}
\safemath{\cvec}{\bmc}
\safemath{\Ddict}{\bD}
\safemath{\Ddictdim}{\ysigdim\times\xsigdim}
\safemath{\dvec}{\bmd}
\safemath{\Ddicttilde}{\widetilde{\bD}}
\safemath{\Bonb}{\bB}
\safemath{\bvec}{\bmb}
\safemath{\Bonbdim}{\ysigdim\times\ysigdim}
\safemath{\noise}{\bmn}
\safemath{\noisedim}{\ysigim}
\safemath{\err}{\bme}
\safemath{\errdim}{\ysigdim}
\safemath{\errset}{\setE}
\safemath{\nerr}{n_e}
\safemath{\delop}{\bP_\errset}
\safemath{\delopc}{\bP_{{\errset}^c}}

\safemath{\cplxi}{\imath}
\safemath{\cplxj}{\jmath}

\safemath{\dict}{\matD}
\safemath{\inputdim}{N}		
\safemath{\outputdim}{M}		
\safemath{\sparsity}{S}	
\safemath{\inputdimA}{{N_a}}	
\safemath{\inputdimB}{{N_b}}	
\safemath{\elemA}{{n_a}}	
\safemath{\elemB}{{n_b}}	
\safemath{\resA}{\matR_a}	
\safemath{\resB}{\matR_b}	
\safemath{\subD}{\matS} 
\safemath{\subA}{\matS_a} 
\safemath{\subB}{\matS_b} 
\safemath{\dicta}{\matA} 	
\safemath{\dictb}{\matB} 	
\safemath{\hollowS}{H}
\safemath{\hollowA}{H_a}
\safemath{\hollowB}{H_b}
\safemath{\cross}{Z}
\safemath{\coh}{\mu_d}			
\safemath{\coha}{\mu_a}			
\safemath{\cohb}{\mu_b}			
\safemath{\mubs}{\nu}	
\safemath{\cohm}{\mu_m} 
\safemath{\dictset}{\setD}	
\safemath{\dictsetp}{\dictset(\coh,\coha,\cohb)}	
\safemath{\dictsetgen}{\dictset_\text{gen}}
\safemath{\dictsetgenp}{\dictsetgen(\coh)}
\safemath{\dictsetonb}{\dictset_\text{onb}}
\safemath{\dictsetonbp}{\dictsetonb(\coh)}

\safemath{\leftside}{U}
\safemath{\rightsideA}{R_a}
\safemath{\rightsideB}{R_b}

\safemath{\indexS}{\setI_S} 

\safemath{\na}{n_a}			
\safemath{\nb}{n_b}			
\safemath{\coeffa}{p_i}	
\safemath{\coeffb}{q_j}	
\safemath{\seta}{\setP}		
\safemath{\setb}{\setQ}     
\safemath{\setw}{\setW}	
\safemath{\setz}{\setZ}	
\safemath{\cola}{\veca}		
\safemath{\colb}{\vecb}		
\safemath{\cold}{\vecd}		
\safemath{\inputvec}{\vecx} 	
\safemath{\error}{\vece}	
\safemath{\noiseout}{\vecz} 	
\safemath{\inputvecel}{x}
\safemath{\inputveca}{\vecx_a}
\safemath{\inputvecb}{\vecx_b}
\safemath{\outputvec}{\vecy}	
\safemath{\lambdamin}{\lambda_{\mathrm{min}}}

\newcommand{\normtwo}[1]{\vecnorm{#1}_2}
\newcommand{\normone}[1]{\vecnorm{#1}_1}
\newcommand{\normzero}[1]{\vecnorm{#1}_0}
\newcommand{\norminf}[1]{\vecnorm{#1}_\infty}

\safemath{\elltwo}{\ell_2}
\safemath{\ellone}{\ell_1}
\safemath{\ellzero}{\ell_0}
\safemath{\ellinf}{\ell_\infty}
\safemath{\ellinftilde}{\ell_{\widetilde\infty}}
\safemath{\licard}{Z(\coh,\coha,\cohb)}
\safemath{\xsol}{\hat{x}}
\safemath{\xbord}{x_b}		
\safemath{\xstat}{x_s}		
\safemath{\xstatLone}{\tilde{x}_s}
\safemath{\order}{\mathcal{O}} 
\safemath{\scales}{\Theta} 
\safemath{\ones}{\mathbf{1}} 
\safemath{\zeroes}{\mathbf{0}} 
\safemath{\thlone}{\kappa(\coh,\cohb)} 
\safemath{\constoneA}{\delta} 
\safemath{\constoneB}{\epsilon} 
\safemath{\nlarge}{L}				   
\safemath{\sumlarge}{S_\nlarge}
\safemath{\maxlarger}{P_\nlarge}	   
\safemath{\Pzero}{\textrm{P0}}	
\safemath{\Pone}{\textrm{P1}}
\safemath{\vecfir}{\vecw}			 
\safemath{\vecsec}{\vecz}
\safemath{\elvecfir}{w}              
\safemath{\elvecsec}{z}				 
\safemath{\nlargefir}{n}
\safemath{\normout}{\gamma}
\safemath{\auxfun}{h}
\safemath{\supp}{\textrm{supp}}

\safemath{\indexa}{\ell}
\safemath{\indexb}{r}
\safemath{\indexc}{i}
\safemath{\indexd}{j}

\safemath{\project}{P}

\IEEEoverridecommandlockouts 

\usepackage{color}

\makeatletter
\g@addto@macro\ps@headings{%
  \def\@oddfoot{\@IEEEfooterstyle\@date\hfil IIP Group Technical Report}%
  \def\@evenfoot{\@IEEEfooterstyle IIP Group Technical Report\hfil\@date}%
}
\g@addto@macro\ps@IEEEtitlepagestyle{%
  \def\@oddfoot{\@IEEEfooterstyle\@date\hfil IIP Group Technical Report}%
  \def\@evenfoot{\@IEEEfooterstyle IIP Group Technical Report\hfil\@date}%
}
\makeatother
\begin{document}

\title{Recovery of Signals with Low Density}

\author{ \phantom{x} \\[-0.3cm]
Christoph Studer\\[0.6cm]
\small \em Integrated Information Processing (IIP) Group Technical Report \\
Department of Information Technology and Electrical Engineering, ETH Z\"urich, Switzerland \\
 e-mail: studer@ethz.ch; website: http://iip.ethz.ch
 \thanks{This report is a revised version of a manuscript first posted to arXiv on July 10, 2015, and subsequently submitted as a short paper to a journal. After the submission ended up in review limbo, the author eventually stopped pursuing it. The present version corrects an error in the original proof of the OMP recovery guarantee, incorporates results omitted from the 2015 version, and updates the discussion and bibliography to reflect both earlier and more recent related work. The revised manuscript is released solely as a technical report, with no further journal submission planned.}
}

\maketitle

\begin{abstract}
Sparse signals (i.e., vectors with a small number of non-zero entries) build the foundation of most kernel (or nullspace) results, uncertainty relations, and recovery guarantees in the sparse signal-processing and compressive-sensing literature. 
In this report, we study a \emph{signal-density measure}, the ratio between the $\ell_1$-norm and the $\ell_\infty$-norm of a vector, which extends the common notion of sparsity to non-sparse signals whose entries' magnitudes decay rapidly. By taking into account such magnitude information, we derive a kernel result and an uncertainty relation that are more general and less restrictive than those based on the $\ell_0$-pseudonorm.
Furthermore, we use this density measure to analyze orthogonal matching pursuit (OMP). We show that OMP provably (i) recovers sparse signals with decaying magnitudes using up to 2$\boldsymbol\times$ more non-zero coefficients than guaranteed by standard, sparsity-based results and (ii) identifies the largest entries of arbitrary signals under a suitable magnitude-decay condition.
\end{abstract}




\section{Introduction}

We consider kernel (or nullspace) results providing conditions for which $\bA\vecx\neq\bZero_{M\times 1}$ and uncertainty relations for pairs of signals (vectors) $\vecx\in\complexset^{N_a}$ and $\vecz\in\complexset^{N_b}$ that satisfy $\bA\vecx=\bB\vecz$, where $\dicta\in\complexset^{M\times N_a}$ and $\dictb\in\complexset^{M\times N_b}$ are dictionaries, i.e., matrices whose columns are normalized to unit $\elltwo$-norm. We furthermore study conditions for which orthogonal matching pursuit (OMP) recovers signals $\vecx$ from an underdetermined system of linear equations $\vecy=\bA\vecx$ with $M<N_a$.

\subsection{Contributions and Outline}

In contrast to existing kernel results, uncertainty relations, and recovery guarantees that have been derived for  \emph{sparse} signals, i.e., for vectors with only a small number of nonzero entries~\cite{gribonval2003,tropp2006,tropp2004,donoho2006a,bruckstein2009sparse,elad2006sparse,elad2002,kuppinger2010a,studer2010,benhaim2010,eladbook2010}, the results developed in this report make use of the following definition (see  \fref{sec:density} for the details).

\begin{defi}[$\delta$-Density] \label{def:densities}
For a non-zero signal \mbox{$\vecx\in\complexset^{N_a}$}, we define its $\delta$-density as follows:
\begin{align} \label{eq:density}
\delta(\vecx)  = {\normone{\vecx}}/{\norminf{\vecx}}.
\end{align}
For an all-zero signal $\vecx = \bZero_{N_a\times1}$, we define $\delta(\vecx)=0$.
\end{defi}

The $\delta$-density has been used before by Tang and Nehorai in~\cite{TN15} to derive computable performance bounds on sparse recovery and coincides with the $q=\infty$ member of the family of $q$-ratio sparsity measures studied more recently~\cite{ZY19qcmsv,ZY19frontiers,ZY21qratio}; see \fref{sec:density} for a detailed discussion of these connections.

Our results in  \fref{sec:kernelresults} show that if $\bA$ has incoherent columns and the signal~$\vecx$ has sufficiently low $\delta$-density, then $\bA\vecx\neq\bZero_{M\times 1}$. 
Furthermore, if the columns among $\bA$ and $\bB$ are incoherent, then two signals~$\vecx$ and $\vecz$ satisfying $\bA\vecx=\bB\vecz$ cannot have low density at the same time. 
In \fref{sec:OMPrecovery}, we show that OMP perfectly recovers a strictly sparse signal $\vecx$ from $\vecy=\bA\vecx$ with up to nearly $1/\coha$ non-zero entries, provided its coefficients decay sufficiently fast; here, $\coha$ is the coherence of~$\dicta$ (see \fref{def:coherence}). This result improves upon standard OMP recovery guarantees by up to $2\times$; we emphasize that a coherence-based improvement of this type was first established by Herzet \emph{et al.} in~\cite{HS14} using consecutive-decay conditions---see \fref{sec:OMPrelated} for a detailed discussion. We further show that OMP identifies the dominant (in magnitude) coefficients of arbitrary signals with low density, under a more restrictive condition. We conclude in \fref{sec:conclusion}.
All proofs are relegated to the appendices.

\subsection{Notation}

Lowercase and uppercase boldface letters stand for column vectors and matrices, respectively. For the matrix~\bM, we denote its adjoint and (left) pseudo-inverse by~$\bM^\herm$ and $\bM^{\dagger}=(\bM^\herm\bM)^{-1}\bM^\herm$, respectively, where the latter requires $\bM$ to have full column rank; $\bI_M$ and $\bZero_{M\times N}$ denote the $M\times M$ identity and $M\times N$ all-zero matrix, respectively; $\bOne_{M\times N}$ denotes the $M\times N$ all-ones matrix, and $\ker(\bM)=\{\vecx:\bM\vecx=\bZero_{M\times1}\}$ is the kernel (or nullspace) of $\bM\in\complexset^{M\times N}$. The $i$th column of the matrix $\bM$ is denoted by $\vecm_i$.
For a vector $\vecx\in\complexset^{N}$ with entries $x_i$, $i=1,\ldots,N$, we define the norms $\normone{\vecx}=\sum_{i=1}^N \abs{x_i}$, $\normtwo{\vecx}=(\sum_{i=1}^N \abs{x_i}^2)^{1/2}$, and $\norminf{\vecx}=\max_{i}\abs{x_i}$; the sparsity $\normzero{\vecx}$ is the number of non-zero entries of $\vecx$.
Sets are denoted by upper-case calligraphic letters; the cardinality of the set \setS is~$\abs{\setS}$.
The matrix $\bM_\setS$ is obtained from \bM by retaining the columns of \bM with indices in~$\setS$; the vector $\bmv_\setS$ is obtained analogously from~\bmv. We define  $[x]^+=\max\{x,0\}$ for $x\in\reals$.


\section{Properties of the \texorpdfstring{$\delta$}{delta}-Density}
\label{sec:density}

The $\delta$-density in \fref{def:densities} enables a finer characterization of important signal properties than the signal sparsity~$\normzero{\vecx}$, which simply counts the number of non-zero entries in $\vecx$.
We next summarize the key properties of the proposed density measure. 

The following result establishes an upper bound on $\delta(\vecx)$; a short proof is given in \fref{app:densitychain}.
\begin{lem}[$\delta$-Density vs.~Signal Sparsity] \label{lem:densitychain}
For any non-zero signal $\vecx\in\complexset^{N_a}$, the $\delta$-density in \fref{def:densities} satisfies
\begin{align} \label{eq:densityrelation}
1\leq \delta(\vecx)  \leq \normzero{\vecx} \leq N_a.
\end{align}
The all-zero signal is excluded, as $\delta(\bZero_{N_a\times1})=0$ by \fref{def:densities}. Equality $\delta(\vecx)=\normzero{\vecx}$ holds if and only if the non-zero entries of $\vecx$ have constant modulus, and $\delta(\vecx)=1$ if and only if $\vecx$ is $1$-sparse.
\end{lem}

The result in \fref{eq:densityrelation} implies that if the signal $\vecx$ is sparse, i.e., $\normzero{\vecx}$ is smaller than its ambient dimension~$N_a$, then it must also have low density (bounded by $\normzero{\vecx}$). 
Furthermore, for signals whose non-zero entries are constant modulus, both the $\delta$-density $\delta(\vecx)$ and the signal sparsity $\normzero{\vecx}$ coincide. 
In contrast, signals with low $\delta$-density need not be sparse.
As an example, consider the following signal class.

\begin{defi}[$\alpha$-Decaying Signal] \label{def:decayingsignal}
We define an $\alpha$-decaying signal as a vector $\vecx\in\reals^{N_a}$ whose entries satisfy $x_i=\alpha^{i-1}$, $i=1,\ldots,N_a$ with $0<\alpha<1$.
\end{defi}

For such signals, the sparsity equals the ambient dimension~$N_a$ as all entries are non-zero; in contrast, the $\delta$-density can be bounded by
\begin{align} \label{eq:alphadensitybound}
\delta(\vecx)\leq\frac{1}{1-\alpha},
\end{align}
which is strictly smaller than $\normzero{\vecx}=N_a$ whenever $\alpha<1-1/N_a$, i.e., whenever the decay is fast relative to the ambient dimension.
Hence, the $\delta$-density does not simply count the number of nonzero entries, but also captures crucial magnitude  information of the signal's non-zero coefficients---all of our results developed in Sections~\ref{sec:kernelresults} and~\ref{sec:OMPrecovery} make use of this particular property.

The $\delta$-density is one of three closely-related density measures. The next result places it within a chain that also contains $\gamma(\vecx)=\normtwo{\vecx}^2/\norminf{\vecx}^2$ and $\sigma(\vecx)=\normone{\vecx}^2/\normtwo{\vecx}^2$; the latter is also known as the numerical (or effective) \emph{sparsity}~\cite{GN11,andersson2014theorem,ELX13}. A short proof is given in \fref{app:densitychain2}.

\begin{lem}[Density Chain] \label{lem:densitychain2}
For any non-zero signal $\vecx\in\complexset^{N_a}$, the density measures $\gamma(\vecx)$, $\delta(\vecx)$, and~$\sigma(\vecx)$ satisfy
\begin{align} \label{eq:densitychain}
1\leq \gamma(\vecx)\leq\delta(\vecx)\leq\sigma(\vecx)\leq\normzero{\vecx},
\end{align}
together with the identity $\delta(\vecx)^2=\gamma(\vecx)\,\sigma(\vecx)$. Furthermore, $\gamma(\vecx)=\delta(\vecx)=\sigma(\vecx)=\normzero{\vecx}$ if and only if the non-zero entries of $\vecx$ have constant modulus.
\end{lem}

In words, $\delta(\vecx)$ is the geometric mean of $\gamma(\vecx)$ and $\sigma(\vecx)$, and all three density measures are sandwiched between~$1$ and the sparsity $\normzero{\vecx}$; this generalizes \fref{lem:densitychain}.

We note that the density measures studied in this report are not new. Tang and Nehorai used the $\sigma$-density~\cite{GN11} and, later, the $\delta$-density~\cite{TN15} as computable matrix-quality measures for $\ell_1$-norm-based sparse recovery. Zhou and Yu subsequently developed the general family of $q$-ratio sparsity measures $s_q(\vecx)=({\normone{\vecx}}/{\|\vecx\|_q})^{q/(q-1)}$ for $1<q\leq\infty$, together with associated recovery guarantees and direct-minimization methods~\cite{ZY19qcmsv,ZY19frontiers,ZY21qratio}. In this terminology, the $\delta$-density and the $\sigma$-density are the cases $\delta(\vecx)=s_\infty(\vecx)$ and $\sigma(\vecx)=s_2(\vecx)$; the $\sigma$-density has also appeared as a sparsity measure in~\cite{andersson2014theorem,ELX13}.
The contribution of this report does not lie in the density measures themselves, but in expressing kernel results, a two-dictionary uncertainty relation, and OMP recovery conditions directly in terms of the coherence and these density measures; to the best of our knowledge, such coherence-explicit results are novel.

While this report primarily focuses on signals with \emph{low} density, the same measures characterize the opposite extreme. The following result concerns \emph{democratic} (i.e., maximally spread) representations~\cite{studer2015democratic} obtained by $\ellinf$-norm minimization over a full-spark frame\footnote{A full-spark frame is a dictionary $\dicta\in\complexset^{M\times N_a}$ for which any $M$ columns of $\dicta$ are linearly independent.}; a short proof is given in \fref{app:democratic}.

\begin{lem}[Density of Democratic Representations] \label{lem:democratic}
Let $\dicta\in\complexset^{M\times N_a}$ be a full-spark frame and $\vecy\in\complexset^M$ a non-zero vector. Then, the minimum-$\ellinf$-norm representation obtained by solving
\begin{align}
\dot\vecx = \arg\min_{\vecx\in\complexset^{N_a}} \norminf{\vecx} \,\textnormal{ subject to } \,\vecy=\dicta\vecx \label{eq:democratic}
\end{align}
satisfies $\sigma(\dot\vecx)\geq \delta(\dot\vecx)\geq\gamma(\dot\vecx)\geq N_a-M+1$.
\end{lem}

Put simply, democratic representations obtained by $\ellinf$-norm minimization provably have \emph{high} density, which is the opposite of the low-density signals studied in the rest of this report.

In what follows, we will mainly focus on the $\delta$-density for reasons that will become clear in \fref{sec:whydelta}.
We first show that the $\delta$-density not only exhibits similarity to the signal sparsity~$\|\vecx\|_0$, but also some fundamental differences.
The following result is a consequence of \fref{def:densities}.
\begin{lem}[Scale Invariance]
The $\delta$-density $\delta(\vecx)$ and $\normzero{\vecx}$ are invariant to scaling by non-zero scalars, i.e., $\delta(c\vecx)=\delta(\vecx)$ and $\normzero{c\vecx}=\normzero{\vecx}$ for every $c\in\complexset$ with $c\neq0$.
\end{lem}

As a consequence, the $\delta$-density and the sparsity $\normzero{\vecx}$ are not norms.  
In contrast to the sparsity $\normzero{\vecx}$, the  $\delta$-density also violates the triangle inequality. A short  proof is given in \fref{app:triangle}.
\begin{lem}[Triangle Inequality] \label{lem:triangle}
In general, the $\delta$-density does not satisfy the triangle inequality.
\end{lem}

This result has immediate negative consequences for uniqueness proofs that rely on the $\delta$-density. Specifically, a common way of establishing uniqueness for sparse signals (see, e.g.,  \cite{eladbook2010,kuppinger2010a,studer2010}) is to consider two signals $\vecx$ and $\vecx'$, and investigate $\bA\vecx=\bA\vecx'$, which implies \mbox{$\bA(\vecx-\vecx')=\bZero_{M\times 1}$}. Thanks to the triangle inequality, the sparsity $\normzero{\vecx-\vecx'}$ can now be bounded by $\normzero{\vecx}+\normzero{\vecx'}$; this approach, however, does not apply to the $\delta$-density, which inhibits the derivation of similar uniqueness guarantees. We therefore provide an alternative recovery condition in \fref{sec:OMPrecovery} for OMP.


\section{Kernel Result and Uncertainty Relation}
\label{sec:kernelresults}

We now develop a kernel (or nullspace) result and uncertainty relation for signals with low $\delta$-density. All our results  make use of the following definitions.  
\begin{defi}[Dictionary Coherence] \label{def:coherence}
Let $\dicta\in\complexset^{M\times N_a}$ be a dictionary. Then
\begin{align} \label{eq:coherence}
\coha=\max_{i\neq j}\, |\veca^\herm_i\veca_j|
\end{align}
is the coherence  of the dictionary $\dicta$. The coherence $\cohb$ of a dictionary $\dictb\in\complexset^{M\times N_b}$ is defined analogously.
\end{defi}

\begin{defi}[Mutual Coherence]
Let $\dicta\in\complexset^{M\times N_a}$ and $\dictb\in\complexset^{M\times N_b}$ be two dictionaries. Then
\begin{align} \label{eq:mutualcoherence}
\cohm=\max_{i,j}\, |\veca^\herm_i\vecb_j|
\end{align}
is the mutual coherence between the dictionaries $\dicta$ and $\dictb$. 
\end{defi}

Note that for an orthonormal basis (ONB)  $\dicta\in\complexset^{M\times M}$, we have \mbox{$\coha=0$}. For a pair of ONBs \mbox{$\dicta,\dictb\in\complexset^{M\times M}$}, we have the following lower bound: \mbox{$\cohm\geq1/\sqrt{M}$}~(see, e.g., \cite{gribonval2003,elad2002}).

\subsection{Kernel Result}

The next lemma is key for establishing our $\delta$-density-based kernel result; the proof is given in \fref{app:normbound1}.

\begin{lem}[Bounds on the $\ell_\infty$ Matrix Norm] \label{lem:normbound1}
Let  $\dicta\!\in\complexset^{M\times N_a}$ be a dictionary with coherence $\coha$ and $\vecx\in\complexset^{N_a}$ a  nonzero signal. Then, the following inequalities hold:
\begin{align} \label{eq:normbound1}
1-\coha(\delta(\vecx)-1) \leq \frac{\norminf{\dicta^\herm\dicta\vecx}}{\norminf{\vecx}}\leq 1+\coha(\delta(\vecx)-1).
\end{align}
\end{lem}

This result resembles lower and upper bounds on the $\ell_\infty$ matrix norm, which corresponds to the maximum absolute row sum of the Gram matrix~$\bA^\herm\bA$.
The bounds in~\fref{eq:normbound1}, however, depend on the $\delta$-density of the signal $\vecx$ and enable us to establish the following kernel (or null space) result; a short proof is given in \fref{app:kernelresult}.

\begin{thm}[$\delta$-Density Kernel Result]\label{thm:kernelresult}
Let $\dicta\in\complexset^{M\times N_a}$ be a dictionary with non-trivial coherence $\coha\neq0$ and $\vecx\in\complexset^{N_a}$ be  a non-zero signal.
If 
\begin{align} \label{eq:kernelcondition}
\delta(\vecx) < 1 + {1}/{\coha},
\end{align}
then $\vecx$ cannot be in the kernel of $\dicta$, i.e., $\dicta\vecx\neq\bZero_{M\times1}$.
\end{thm}

This result implies that non-sparse signals (e.g., with up to $N_a$ non-zero entries) having sufficiently low density \emph{cannot} be in the kernel of an incoherent dictionary. To see this, consider an $\alpha$-decaying signal with
\begin{align} \label{eq:alphakernelcondition}
0<\alpha<1-(1+1/\coha)^{-1}.
\end{align}
For such signals, we have $\normzero{\vecx}=N_a$, whereas the $\delta$-density based condition in \fref{eq:kernelcondition} is always met. Hence, suitably defined density measures enable us to establish the important fact that not only sufficiently sparse signals cannot be in the  kernel of incoherent dictionaries but also certain non-sparse signals with sufficiently low $\delta$-density.
\fref{thm:kernelresult} also recovers existing nullspace conditions as a special case. From $\delta(\vecx)\leq\normzero{\vecx}$ we obtain the well-known (and more restrictive) condition $\normzero{\vecx}<1+{1}/{\coha}$; see, e.g., \cite{eladbook2010,gribonval2003,bruckstein2009sparse} for more details. 

\subsection{Restricted Isometry Property}
\label{sec:RIP}

\fref{lem:normbound1} bounds the $\ellinf$-action of the Gram matrix $\dicta^\herm\dicta$ in terms of the $\delta$-density. We now provide a related result that bounds the energy $\normtwo{\dicta\vecx}^2$ in terms of the $\sigma$-density; the proof is given in \fref{app:normbound2}.

\begin{lem}[$\sigma$-Density Restricted Isometry] \label{lem:normbound2}
Let $\dicta\in\complexset^{M\times N_a}$ be a dictionary with coherence $\coha$ and $\vecx\in\complexset^{N_a}$ a non-zero signal. Then, the following inequalities hold:
\begin{align} \label{eq:dictbound2}
1-\coha(\sigma(\vecx)-1) \leq \frac{\normtwo{\dicta\vecx}^2}{\normtwo{\vecx}^2}\leq 1+\coha(\sigma(\vecx)-1).
\end{align}
\end{lem}

This result is a coherence-based restricted isometry property (RIP)~\cite{candes2008a}, stated for the $\sigma$-density rather than the signal sparsity.\footnote{Strictly speaking, \fref{lem:normbound2} is a pointwise, signal-dependent energy bound rather than a RIP in the usual uniform sense. However, since the bounds in \fref{eq:dictbound2} are monotone in $\sigma(\vecx)$, they immediately imply the uniform RIP-type statement $1-\coha(s-1)\leq\normtwo{\dicta\vecx}^2/\normtwo{\vecx}^2\leq1+\coha(s-1)$ over the set $\{\vecx\neq\bZero_{N_a\times1}:\sigma(\vecx)\leq s\}$.}
Since $\sigma(\vecx)\leq\normzero{\vecx}$ by \fref{lem:densitychain2}, it recovers---and is generally tighter than---the classical coherence-based bound $1\pm\coha(\normzero{\vecx}-1)$ obtained for $\normzero{\vecx}$-sparse signals via Ger\v{s}gorin's disc theorem.
As an immediate consequence, \fref{lem:normbound2} yields a second kernel result; a short proof is given in \fref{app:sigmakernel}.

\begin{cor}[$\sigma$-Density Kernel Result] \label{cor:sigmakernel}
Let $\dicta\in\complexset^{M\times N_a}$ have non-trivial coherence $\coha\neq0$ and let $\vecx\in\complexset^{N_a}$ be a non-zero signal. If $\sigma(\vecx)<1+1/\coha$, then $\dicta\vecx\neq\bZero_{M\times1}$.
\end{cor}

\subsection{Why the \texorpdfstring{$\delta$}{delta}-Density?}
\label{sec:whydelta}

We have established two coherence-based kernel results, one for the $\delta$-density (\fref{thm:kernelresult}) and one for the~$\sigma$-density (\fref{cor:sigmakernel}). We can now place them---along with the $\gamma$-density---in context and explain why $\delta$ is our density measure of choice.

The density chain \fref{eq:densitychain} orders the three measures as $\gamma(\vecx)\leq\delta(\vecx)\leq\sigma(\vecx)$. By \fref{thm:kernelresult} and \fref{cor:sigmakernel}, both the $\delta$-density and the $\sigma$-density admit a kernel result of the same form, namely that a density lower than $1+1/\coha$ implies $\dicta\vecx\neq\bZero_{M\times1}$; since $\delta(\vecx)\leq\sigma(\vecx)$, the $\delta$-density-based condition \fref{eq:kernelcondition} is the less restrictive of the two.
It is therefore natural to ask whether the even smaller $\gamma$-density admits a kernel result of the same form, i.e., with a threshold that depends only on the coherence $\coha$ but \emph{not} on the ambient dimension~$N_a$. The answer is negative.
To see this, normalize $\norminf{\vecx}=1$ and set $y_i=\abs{x_i}$, so that $\delta(\vecx)-1=\sum_{i\neq \hat k} y_i$ and $\gamma(\vecx)-1=\sum_{i\neq \hat k} y_i^2$, where $\hat k$ is the index of a largest-magnitude entry of~$\vecx$.
By the proof of \fref{thm:kernelresult}, every kernel vector must satisfy $\delta(\vecx)\geq1+1/\coha$, i.e., $\sum_{i\neq\hat k} y_i\geq 1/\coha$, which is an $\ell_1$-norm-type condition that the $\gamma$-density cannot detect: a low-amplitude tail consisting of many entries can carry the required $\ell_1$-norm while contributing negligibly to the sum $\sum_{i\neq\hat k} y_i^2$. Concretely, a ``spike-plus-uniform-tail'' vector with $y_i=1/(\coha(N_a-1))$ for all $i\neq\hat k$ satisfies $\delta(\vecx)=1+1/\coha$, whereas $\gamma(\vecx)=1+1/(\coha^2(N_a-1))\to1$ as $N_a\to\infty$.
In \fref{app:gammaconstruction}, we provide an explicit construction showing that such spike-plus-uniform-tail vectors are not merely numerically admissible, but are actual kernel vectors of dictionaries with prescribed coherence $\coha$; hence, no dimension-independent $\gamma$-density threshold larger than one can exclude kernel membership.
Thus, the strongest possible $\gamma$-density-based statement is necessarily dimension-dependent: the Cauchy--Schwarz inequality gives $\gamma(\vecx)-1=\sum_{i\neq\hat k} y_i^2\geq\big(\sum_{i\neq\hat k} y_i\big)^2/(N_a-1)=(\delta(\vecx)-1)^2/(N_a-1)$, so every kernel vector satisfies
\begin{align} \label{eq:gammakernelbound}
\gamma(\vecx)\geq 1+\frac{(\delta(\vecx)-1)^2}{N_a-1}\geq 1+\frac{1}{\coha^2(N_a-1)},
\end{align}
which is not meaningful for large $N_a$ and is attained with equality---in both inequalities---by the kernel vector constructed in \fref{app:gammaconstruction}, which has a uniform tail and exactly satisfies $\delta(\vecx)=1+1/\coha$.

In summary, among the three considered density measures, $\gamma$, $\delta$, and $\sigma$, the $\delta$-density is the smallest (and least restrictive) measure for which a dimension-free, coherence-based kernel result exists.


\subsection{Uncertainty Relation}
\label{sec:uncertaintyrelations}

We next provide a $\delta$-density-based uncertainty relation for pairs of dictionaries; the proof is given in \fref{app:uncertainty}.
\begin{thm}[Uncertainty Relation] \label{thm:uncertainty}
Let $\vecx\in\complexset^{N_a}$ and $\vecz\in\complexset^{N_b}$ be non-zero vectors satisfying $\dicta\vecx=\dictb\vecz$, and let $\dicta\in\complexset^{M\times N_a}$ and $\dictb\in\complexset^{M\times N_b}$ be two dictionaries. Then, the following inequality holds
\begin{align} \label{eq:uncertaintyprinciple}
[1\!-\!\coha(\delta(\vecx)\!-\!1)]^+[1\!-\!\cohb(\delta(\vecz)\!-\!1)]^+\!\leq \delta(\vecx)\delta(\vecz)\cohm^2.
\end{align}
\end{thm}

This uncertainty relation generalizes the result in \cite{studer2010,kuppinger2010a}
 for sparse signals to the case of signals characterized by the $\delta$-density. As a consequence of \fref{thm:uncertainty}, we have the following result.

\begin{cor}[Uncertainty Relation for Pairs of ONBs]\label{cor:pairONBresult}
For a pair of ONBs $\dicta$ and $\dictb$, the following uncertainty relation holds:
\begin{align} \label{eq:ONBuncertainty}
{1}/{\cohm^2}\leq \delta(\vecx)\delta(\vecz).
\end{align}
\end{cor}

For maximally-incoherent ONBs (e.g., $\bA$ being a Hadamard basis and $\bB$ the identity), we have \mbox{$\cohm=1/\sqrt{M}$} \cite{gribonval2003}. In this case, the uncertainty relation in \fref{cor:pairONBresult} leads to  $M\leq\delta(\vecx)\delta(\vecz)$, which generalizes the well-known uncertainty relation $M\leq\normzero{\vecx}\normzero{\vecz}$  by Donoho and Stark~\cite{donoho1989}.
As a consequence, for a pair of incoherent ONBs, a signal with low density in basis $\dicta$ cannot have low-density in basis $\dictb$, and vice versa. Again, we emphasize that  sparsity is not the key property that is required to establish such uncertainty relations, but rather a suitably-chosen measure of signal density.


\section{Recovery of Signals with Low \texorpdfstring{$\delta$}{delta}-Density}
\label{sec:OMPrecovery}

We now investigate signal recovery via OMP and present two results. The first result, \fref{thm:OMPrecoveryA}, targets \emph{strictly sparse} signals whose non-zero entries decay in magnitude; this result shows that OMP recovers such signals perfectly, even when their sparsity exceeds the classical coherence threshold by up to a factor of~$2\times$. The second result, \fref{thm:OMPrecoveryB}, drops the sparsity assumption and guarantees, under a more restrictive condition, that OMP identifies the largest entries of an \emph{arbitrary} signal. Both results build on the $\delta$-density and make the role of magnitude decay explicit.

\subsection{Orthogonal Matching Pursuit (OMP)}
\label{sec:OMPalgorithm}

OMP~\cite{Pati1993,tropp2010a,eladbook2010} is an iterative method  used to recover sparse vectors $\vecx$ from $\vecy=\dicta\vecx$. After initializing the residual $\vecr^{(0)}=\vecy$ and an empty support set $\setS^{(0)}=\varnothing$, OMP selects a column of the dictionary $\bA$ in every iteration $t=1,\ldots,t_\textnormal{max}$ according to 
\begin{align} \label{eq:OMPselectioncriterion}
\hat k^{(t)} = \arg\max_{i\in\setR^{(t-1)}} \,\abs{\veca^\herm_i\vecr^{(t-1)}}.
\end{align}
Here, the set $\setR^{(t-1)}=\{1,\ldots,N_a\}\backslash\setS^{(t-1)}$ contains all remaining indices that are not (yet) in the support set $\setS^{(t-1)}$. 
Then, the index~$\hat k^{(t)}$ in \fref{eq:OMPselectioncriterion} is added to the new support set $\setS^{(t)}=\setS^{(t-1)}\cup \{\hat k^{(t)}\}$ and a new residual is computed as
\begin{align} \label{eq:OMPLSestimate}
\vecr^{(t)} = \vecy - \bA_{\setS^{(t)}}\hat\vecx_{\setS^{(t)}} = (\bI_M - \bA_{\setS^{(t)}}\bA^\dagger_{\setS^{(t)}})\vecy,
\end{align}
where $\hat\vecx_{\setS^{(t)}}$ is the least-squares estimate of the signal's coefficients on the current support set $\setS^{(t)}$.
The above iterative procedure is repeated until a (predefined) number of iterations~$t_\textnormal{max}$ is reached. The algorithm's outputs are the least-squares estimate $\hat\vecx_{\setS^{(t_\textnormal{max})}}$ and the support-set estimate $\setS^{(t_\textnormal{max})}$.

\subsection{Recovery Guarantees}
\label{sec:OMPguarantees}

We start with strictly-sparse signals whose non-zero entries have decaying magnitudes. For such signals, OMP recovers the support---and therefore the entire signal---exactly, even when the sparsity exceeds the classical OMP recovery threshold~\cite{tropp2004,eladbook2010}
\begin{align} \label{eq:classicalOMPcondition}
\normzero{\vecx} < \textstyle \frac{1}{2}\!\left(1+{1}/{\coha}\right).
\end{align}
The proof is given in \fref{app:OMPproofA}.

\begin{thm}[Recovery of Strictly-Sparse Signals]\label{thm:OMPrecoveryA}
Let $\dicta$ have non-trivial coherence $\coha\neq0$, and let $\vecx\in\complexset^{N_a}$ be a non-zero signal with support $\setX=\{i:x_i\neq0\}$ and strict sparsity $s=\normzero{\vecx}$. Run OMP with $t_\textnormal{max}=s$ iterations. If, in every iteration $t=1,\ldots,s$, the following condition holds
\begin{align} \label{eq:OMPconditionA}
\delta\big(\vecx_{\setX\backslash\setS^{(t-1)}}\big) < \tfrac{1}{2}\big(1+{1}/{\coha}-(t-1)\big), 
\end{align}
then OMP selects an atom from the support $\setX$ in every iteration. Consequently, $\setS^{(s)}=\setX$ and OMP recovers $\vecx$ exactly.
\end{thm}

Condition~\fref{eq:OMPconditionA} constrains the $\delta$-density of the (as yet) unrecovered part of the signal; a convenient \emph{a priori} sufficient condition is that~\fref{eq:OMPconditionA} hold for every proper subset $\setS\subsetneq\setX$ with $t-1=|\setS|$. 
Moreover, since $\delta(\vecx_{\setX\backslash\setS^{(t-1)}})\geq1$, as long as not all of the set~$\setX$ has been selected, condition \fref{eq:OMPconditionA} implies $1<\frac{1}{2}(1+1/\coha-(t-1))$ and, hence, $t<1/\coha$. By Ger\v{s}gorin's disc theorem, the smallest eigenvalue of $\bA^\herm_{\setS^{(t)}}\bA_{\setS^{(t)}}$ is therefore at least $1-\coha(t-1)>1-\coha(1/\coha-1)=\coha>0$, so the pseudo-inverse~$\bA^\dagger_{\setS^{(t)}}$ in~\fref{eq:OMPLSestimate} exists in every iteration.

Our second result makes no sparsity assumption and  applies to \emph{arbitrary} signals, including those with up to $N_a$ non-zero entries. This results in a more restrictive condition that depends explicitly on the gap between the two largest remaining magnitudes. The proof is given in \fref{app:OMPproofB}.

\begin{thm}[Identification of the Largest Entries]\label{thm:OMPrecoveryB}
Let $\dicta$ have non-trivial coherence $\coha\neq0$ and let $\vecx\in\complexset^{N_a}$ be a non-zero signal. Suppose that the maximum number of OMP iterations satisfies $t_\textnormal{max}\leq\normzero{\vecx}$ and
\begin{align} \label{eq:OMPfirstcondition}
t_\textnormal{max} < 1+ {1}/{\coha},
\end{align}
and, in every iteration $t=1,\ldots,t_\textnormal{max}$, the following condition holds: 
\begin{align} \label{eq:OMPconditionB}
\delta\big(\vecx_{\setR^{(t-1)}}\big) < \tfrac{1}{2}\big(1+{1}/{\coha}-(t-1)\big)\Big(1-\tfrac{\rho^{(t)}}{1+\coha}\Big).
\end{align}
Here, $\rho^{(t)}=m_2^{(t)}/m_1^{(t)}\in[0,1)$, where $m_1^{(t)}=\norminf{\vecx_{\setR^{(t-1)}}}$ denotes the largest magnitude among the remaining coefficients and $m_2^{(t)}$ denotes the largest remaining magnitude that is strictly smaller than $m_1^{(t)}$ (with $m_2^{(t)}=0$ if no such entry exists). 
Then, OMP selects an atom associated with (one of) the largest coefficient(s) in $\vecx_{\setR^{(t-1)}}$ in iteration $t$, and the support set $\setS^{(t)}$ indexes the $t$ largest (in magnitude) entries of $\vecx$. If, in addition, $\normzero{\vecx}=t_\textnormal{max}$, then OMP recovers $\vecx$ exactly.
\end{thm}

The condition $t_\textnormal{max}\leq\normzero{\vecx}$ ensures that at least one non-zero coefficient remains in every iteration, so that $m_1^{(t)}>0$ and $\rho^{(t)}$ is well-defined; without it, the notion of ``largest remaining entries'' becomes meaningless once all non-zero entries have been selected. The condition~\fref{eq:OMPfirstcondition} ensures that the pseudo-inverse~$\bA^\dagger_{\setS^{(t)}}$ in~\fref{eq:OMPLSestimate} exists in every iteration. The condition~\fref{eq:OMPconditionB} is the corrected counterpart of~\fref{eq:OMPconditionA} for non-sparse signals: the correction factor $(1-\rho^{(t)}/(1+\coha))\le1$ accounts for the fact that, when $\vecx$ is not sparse, the residual retains energy along the not-yet-selected coefficients, which can mislead the selection criterion in~\fref{eq:OMPselectioncriterion} unless the largest magnitude is sufficiently separated from the next-largest entry.

We conclude by noting that, although conditions \fref{eq:OMPconditionA} and \fref{eq:OMPconditionB} formally involve the support sets selected by OMP, both can be verified a priori: for \fref{thm:OMPrecoveryA} via the all-subsets condition discussed after the theorem and for \fref{thm:OMPrecoveryB} directly, because the theorem inductively guarantees that $\setS^{(t-1)}$ consists of the $t-1$ largest-magnitude entries of $\vecx$, so that $\delta(\vecx_{\setR^{(t-1)}})$, $m_1^{(t)}$, $m_2^{(t)}$, and $\rho^{(t)}$ are all determined by the sorted magnitude profile of~$\vecx$ (ties do not affect these quantities). Both theorems are therefore a priori guarantees rather than trajectory-conditional statements.

\subsection{Discussion}
\label{sec:OMPdiscussion}

\subsubsection{Constant-Modulus Signals} 
We emphasize that for signals whose non-zero entries have constant modulus, both guarantees collapse to the well-known OMP recovery threshold in \fref{eq:classicalOMPcondition}.
Indeed, for such signals $\delta(\vecx_{\setR^{(t-1)}})=\normzero{\vecx}-(t-1)$ and $\rho^{(t)}=0$. Setting $t_\textnormal{max}=\normzero{\vecx}$, both~\fref{eq:OMPconditionA} and~\fref{eq:OMPconditionB} are then met in all iterations if and only if \fref{eq:classicalOMPcondition} holds; the binding case being the first iteration $t=1$. 
This shows (i) that our conditions do not contradict existing recovery guarantees for OMP and (ii) that sparse constant-modulus signals are worst-case signals from an OMP perspective---this aspect is well-known and can also be observed via numerical simulations.

\subsubsection{Benefit of Magnitude Decay} 
In stark contrast, signals with rapidly decaying magnitudes have low $\delta$-density and are recovered well beyond the classical recovery threshold in~\fref{eq:classicalOMPcondition}. Consider again an $\alpha$-decaying signal (\fref{def:decayingsignal}), truncated to its $s$ largest entries, and note that every sub-vector $\vecx_\setP$ with $\setP\subseteq\{1,\ldots,s\}$ satisfies $\delta(\vecx_\setP) \leq (1-\alpha)^{-1}$, since its entries are dominated by a geometric series with ratio~$\alpha$ whose largest term is the entry with the smallest index in $\setP$. Since this bound holds uniformly, the binding case of~\fref{eq:OMPconditionA} is the last iteration ($t=s$), and \fref{thm:OMPrecoveryA} guarantees exact recovery whenever
\begin{align} \label{eq:alphaOMP}
s< {1}/{\coha}+2-{2}/(1-\alpha).
\end{align}
Letting $\alpha\to0$, which corresponds to fast coefficient decay, this becomes $s<1/\coha$, which is roughly $2\times$ less restrictive than the standard condition in~\fref{eq:classicalOMPcondition}. Hence, OMP perfectly recovers a truncated $\alpha$-decaying signal with up to nearly $1/\coha$ non-zero entries. We emphasize that decay-based guarantees approaching the $1/\coha$ threshold were first established by Herzet \emph{et al.}~\cite{HS14}; see \fref{sec:OMPrelated} for a detailed comparison.

\subsubsection{Sparse vs.~Non-Sparse Recovery} 
Theorems \ref{thm:OMPrecoveryA} and \ref{thm:OMPrecoveryB} trade generality for stringency. \fref{thm:OMPrecoveryA} assumes exact sparsity and only asks OMP to recover the support (in any order); this is precisely what yields the $2\times$ gain in~\fref{eq:alphaOMP}. \fref{thm:OMPrecoveryB} handles arbitrary, possibly non-sparse signals and additionally certifies that OMP visits the entries in order of decreasing magnitude, but its condition carries the correction factor $(1-\rho^{(t)}/(1+\coha))\le1$ and is therefore more restrictive. 
As the decay becomes infinitely fast ($\rho^{(t)}\to0$), this factor tends to one and condition~\fref{eq:OMPconditionB} approaches~\fref{eq:OMPconditionA}.

\subsection{Related Results}
\label{sec:OMPrelated}

Coherence-based OMP guarantees that exploit coefficient decay predate the 2015 version of this report. In particular, Herzet \emph{et al.}~\cite{HS14} established exact support-recovery conditions for OMP (and orthogonal least squares) that embed information about the decay of the sorted non-zero magnitudes of strictly sparse signals: under sufficiently strong decay, the standard condition \fref{eq:classicalOMPcondition} relaxes toward $\normzero{\vecx}<1/\coha$, both in the noiseless and in the bounded-noise setting, and they prove that the $1/\coha$ limit is the tightest guarantee achievable from the coherence alone. The limiting threshold $s<1/\coha$ obtained from \fref{thm:OMPrecoveryA} via \fref{eq:alphaOMP}, including the ``up to $2\times$'' improvement claim, is therefore not new. 
What distinguishes our results is the route taken and the form of the conditions: (i) condition \fref{eq:OMPconditionA} constrains a single aggregate quantity---the $\delta$-density of the not-yet-recovered part of the signal---rather than the decay between consecutive sorted magnitudes as in~\cite{HS14}; the two families of conditions are parameterized differently and are, in general, not directly comparable; (ii) the derivation is elementary and self-contained, building only on \fref{lem:normbound1}; and (iii) \fref{thm:OMPrecoveryB} covers arbitrary, possibly non-sparse signals and certifies that OMP selects entries in order of decreasing magnitude, a regime not considered in~\cite{HS14}. Furthermore, the tightness result of~\cite{HS14} implies that the limiting threshold $s<1/\coha$ in \fref{eq:alphaOMP} cannot be improved by any coherence-only argument.

A related recovery result for OMP applied to signals with fast-decaying entries was developed in~\cite{DW10}; that result, however, relies on the RIP (cf.~\fref{sec:RIP}), which is a common way of characterizing measurement matrices in compressive sensing~\cite{candes2008a} but whose parameters are hard to compute in practice. More recently, Wen \emph{et al.}~\cite{WZY20} carried out a signal-dependent performance analysis of OMP for random measurement matrices in which prior information on the non-zero entries is condensed into an upper bound on $\normone{\vecx}^2/\normtwo{\vecx}^2$; this is precisely the $\sigma$-density in \fref{lem:densitychain2}. Furthermore, their probabilistic analysis is complementary to the deterministic, coherence-based conditions derived here.

Finally, the density measures studied in this report have also been used directly as optimization objectives for sparse signal recovery: Wang and Ma~\cite{WANG2023109104} developed iterative shrinkage-thresholding algorithms that minimize the $\ell_1$-over-$\ell_\infty$-norm ratio, i.e., the $\delta$-density, and methods that minimize the $\ell_1$-over-$\ell_2$-norm ratio, which is related to minimizing the $\sigma$-density~\cite{ZY19frontiers,ZY21qratio}. 
Moreover, analytical recovery guarantees for basis pursuit, the Dantzig selector, and the least absolute shrinkage and selection operator (LASSO) that build on $q$-ratio constrained minimal singular values were established in~\cite{TN15,ZY19qcmsv}. Such density-minimization approaches have been shown to empirically outperform conventional $\ell_1$-norm minimization in certain settings, and are complementary to the analytical recovery guarantees developed in this report.


\section{Conclusions}
\label{sec:conclusion}

We have developed kernel results and an uncertainty relation for signals with low density, generalizing existing results for sparse signals to non-sparse signals of sufficiently low density. Furthermore, we have shown that OMP perfectly recovers sparse signals with decaying magnitudes---identifying up to nearly $1/\coha$ non-zero coefficients, roughly twice as many as guaranteed by classical coherence-based conditions, in line with the decay-aware guarantees of~\cite{HS14}---and that it identifies the largest entries of arbitrary signals under a suitable decay condition.
Our results highlight that an appropriately chosen \emph{density measure}, rather than sparsity alone, is key for establishing kernel results, uncertainty relations, and recovery guarantees.

There are many avenues for future work. First, by modifying the proofs of \fref{thm:OMPrecoveryA} and \fref{thm:OMPrecoveryB}, one can obtain recovery conditions in the presence of bounded measurement noise $\vecy=\dicta\vecx+\vecn$ with $\normtwo{\vecn}\leq\varepsilon$ using techniques developed in~\cite{donoho2006a}. 
Second, our proofs extend readily to block-sparse signals~\cite{kuppinger2009block} with suitably defined density measures. 
Third, it would be interesting to sharpen the decay-gap condition~\fref{eq:OMPconditionB} and to investigate density measures beyond the considered $\gamma$, $\delta$, and~$\sigma$ densities. 


\section*{Acknowledgments}

The author thanks Tom Goldstein and Helmut B\"olcskei for insightful discussions, and Ramina Ghods and Charles Jeon for comments on an early version of this report. 

  
\appendices


\section{Basic Properties}

\subsection[Proof of the Density-vs-Sparsity Lemma]{Proof of \fref{lem:densitychain}}
\label{app:densitychain}

From \fref{def:densities}, we have 
\begin{align} \label{eq:densitystuff1}
\delta(\vecx) = \frac{\sum_{i \in \setX}|x_i|}{\max_k |x_k|} = \sum_{i\in\setX} 
\frac{|x_i|}{|x_{\hat{k}}|} \leq |\setX|=\normzero{\vecx},
\end{align}
where $\setX$ is the support set of the signal $\vecx$ and $\hat{k}$ is the index associated to (one of) the largest coefficient(s) of $\vecx$ in terms of magnitude. The inequality in \fref{eq:densitystuff1} follows from the fact that $|x_i|/|x_{\hat{k}}|\leq1$, for $i\in\setX$. Equality holds if and only if $|x_i|/|x_{\hat{k}}|=1$, $i\in\setX$, which implies that  only signals whose nonzero entries have constant modulus satisfy $\delta(\vecx) = \normzero{\vecx}$. The lower bound $\delta(\vecx)\geq1$ holds because the sum in \fref{eq:densitystuff1} contains the term $|x_{\hat{k}}|/|x_{\hat{k}}|=1$, possibly along with non-negative summands, with equality if and only if $\vecx$ has a single non-zero entry.

\subsection[Proof of the Density-Chain Lemma]{Proof of \fref{lem:densitychain2}}
\label{app:densitychain2}

The identity $\delta(\vecx)^2=\gamma(\vecx)\sigma(\vecx)$ follows directly from 
\begin{align}
\gamma(\vecx)\sigma(\vecx)=\frac{\normtwo{\vecx}^2}{\norminf{\vecx}^2}\frac{\normone{\vecx}^2}{\normtwo{\vecx}^2}=\frac{\normone{\vecx}^2}{\norminf{\vecx}^2}=\delta(\vecx)^2.
\end{align}
For the chain \fref{eq:densitychain}, we first note $\gamma(\vecx)\geq1$ because $\normtwo{\vecx}^2\geq\norminf{\vecx}^2$. Next, since $|x_i|^2\leq|x_i|\norminf{\vecx}$ for all~$i$, summing yields $\normtwo{\vecx}^2\leq\normone{\vecx}\norminf{\vecx}$, and therefore
\begin{align}
\delta(\vecx)=\frac{\normone{\vecx}}{\norminf{\vecx}}\leq\frac{\normone{\vecx}^2}{\normtwo{\vecx}^2}=\sigma(\vecx).
\end{align}
Combined with the identity $\delta(\vecx)^2=\gamma(\vecx)\sigma(\vecx)$, this gives $\gamma(\vecx)=\delta(\vecx)^2/\sigma(\vecx)\leq\delta(\vecx)$. Finally, the Cauchy--Schwarz inequality $\normone{\vecx}^2\leq\normzero{\vecx}\normtwo{\vecx}^2$ gives $\sigma(\vecx)\leq\normzero{\vecx}$.
Equality $\normtwo{\vecx}^2=\normone{\vecx}\norminf{\vecx}$ (i.e., $\delta(\vecx)=\sigma(\vecx)$) requires $|x_i|(\norminf{\vecx}-|x_i|)=0$ for all~$i$, i.e., $|x_i|\in\{0,\norminf{\vecx}\}$; equality (i.e., $\sigma(\vecx)=\normzero{\vecx}$) requires the same. Together with \fref{lem:densitychain}, all three quantities coincide with $\normzero{\vecx}$ if and only if the non-zero entries of $\vecx$ have constant modulus.

\subsection[Proof of the Democratic-Representations Lemma]{Proof of \fref{lem:democratic}}
\label{app:democratic}

Since $\dicta$ is a full-spark frame, every representation $\dot\vecx$ with $\vecy=\dicta\dot\vecx$ obtained by $\ellinf$-norm minimization satisfies $\normtwo{\dot\vecx}^2\geq\norminf{\dot\vecx}^2(N_a-M+1)$~\cite[Thm.~3]{studer2015democratic}, which is $\gamma(\dot\vecx)\geq N_a-M+1$ by definition of the $\gamma$-density. The claim $\sigma(\dot\vecx)\geq\delta(\dot\vecx)\geq\gamma(\dot\vecx)$ then follows from the density chain \fref{eq:densitychain}.

\subsection[Proof of the Triangle-Inequality Lemma]{Proof of \fref{lem:triangle}}
\label{app:triangle}

The proof follows from a counterexample. Consider two signals $\vecx\in\reals^{N_a}$ and $\vecz\in\reals^{N_a}$ with entries  $x_i=-\alpha^{i-1}$ and $z_i=\alpha^{i-1}+\varepsilon$ for $i=1,\ldots,N_a$ with $0<\alpha< 1$ and $\varepsilon>0$. Both vectors have  the following $\delta$-densities:
\begin{align} \label{eq:triangledensities}
\delta(\vecx) =  \frac{1-\alpha^{N_a}}{1-\alpha} \quad \text{and} \quad \delta(\vecz) = \frac{1}{1+\varepsilon}\!\left(\frac{1-\alpha^{N_a}}{1-\alpha}+N_a\varepsilon\right)\!,
\end{align}
both of which approach $1$ as $\alpha\to0$ and $\varepsilon\to0$.
However, all entries of the sum of both signals are equal to $\varepsilon$, i.e.,
$\delta(\vecx+\vecz) =  \delta(\varepsilon\bOne_{N_a\times 1}) = N_a$.
Hence, for sufficiently small values of $\alpha$ and $\varepsilon$, and an ambient dimension $N_a\geq3$, we have $\delta(\vecx+\vecz)=N_a>\delta(\vecx)+\delta(\vecz)$, so these signals violate the triangle inequality.

\section{Proof of Kernel Result, Restricted Isometry Property, and Uncertainty Relation}

\subsection[Proof of the Matrix-Norm-Bound Lemma]{Proof of \fref{lem:normbound1}}
\label{app:normbound1}

We start with the lower bound in \fref{eq:normbound1}. Since~$\dicta$ is a dictionary with $\veca^\herm_i\veca_i=1$, $\forall i$, we obtain the following lower bound:
\begin{align}
\norminf{\dicta^\herm\dicta\vecx} 
& = \textstyle  \max_i\,\abs{\veca^\herm_i\veca_i x_i + \sum_{j,i\neq j} \veca^\herm_i\veca_jx_j}   \textstyle  \geq \max_i\!\left\{|x_i| - \sum_{j,i\neq j} \abs{\veca^\herm_i\veca_j}\abs{x_j}\right\},
\end{align}
where the last step follows from the reverse and regular triangle inequalities, respectively. Using the definition of the coherence $\coha$, we obtain
\begin{align} \label{eq:neatreformulation}
\norminf{\dicta^\herm\dicta\vecx}  & \geq \textstyle \max_i\!\left\{\abs{x_i} - \sum_{j,i\neq j} \coha |x_j|\right\} 
 = \textstyle \max_i\!\left\{ |x_i|(1+\coha) - \coha\sum_{j}\abs{x_j} \right\} \notag \\
 &=  \norminf{\vecx}(1+\coha) - \coha\normone{\vecx}.
\end{align}
By excluding the case $\norminf{\vecx}=0$ (implying $\vecx\neq\bZero_{N_a\times1}$), we finally get the lower bound in \fref{eq:normbound1}.
The upper bound in \fref{eq:normbound1} is obtained similarly to \fref{eq:neatreformulation} and is given by
\begin{align} \label{eq:isometryupperbound}
\norminf{\dicta^\herm\dicta\vecx}  
 \leq \textstyle\max_i\!\left\{|x_i| + \sum_{j,i\neq j} \abs{\veca^\herm_i\veca_j}\abs{x_j}\right\}
 \leq \norminf{\vecx}(1-\coha) + \coha\normone{\vecx}.
\end{align}
By excluding $\norminf{\vecx}=0$, we obtain the upper bound in \fref{eq:normbound1}. 
Note that the last step in \fref{eq:isometryupperbound} uses $\max_i\{|x_i|(1-\coha)+\coha\normone{\vecx}\}=\norminf{\vecx}(1-\coha)+\coha\normone{\vecx}$, which requires $\coha\leq1$; this always holds by the Cauchy--Schwarz inequality, as the columns of $\dicta$ have unit $\elltwo$-norm.

\subsection[Proof of the Delta-Density Kernel Result]{Proof of \fref{thm:kernelresult}}
\label{app:kernelresult}
Assume that $\vecx\neq\bZero_{N_a\times1}$ and $\vecx\in\ker(\dicta)$. From the left-hand side (LHS) of \fref{eq:normbound1}, it follows that
\begin{align} \label{eq:kernelproofstep1}
0 \geq (1-\coha(\delta(\vecx)-1)) \norminf{\vecx}
\end{align}
and hence, we have
\begin{align} \label{eq:kernelproofstep2}
\delta(\vecx)\geq1+{1}/{\coha}.
\end{align}
As a consequence,  if $\delta(\vecx)<1+1/\coha$, then $\vecx\notin\ker(\dicta)$.

\subsection[Proof of the Sigma-Density Restricted-Isometry Lemma]{Proof of \fref{lem:normbound2}}
\label{app:normbound2}

We start with the lower bound in \fref{eq:dictbound2}. Expanding the squared $\elltwo$-norm gives
\begin{align} \label{eq:dictboundstep1}
\normtwo{\dicta\vecx}^2 = \vecx^\herm\dicta^\herm\dicta\vecx = \normtwo{\vecx}^2 + \vecx^\herm\bE\vecx,
\end{align}
where $\bE=\dicta^\herm\dicta-\bI_{N_a}$ is the hollow Gram matrix. Writing $\bD=\coha\bI_{N_a}$, we obtain
\begin{align} \label{eq:dictboundstep2}
\normtwo{\dicta\vecx}^2 = \normtwo{\vecx}^2(1+\coha) + \vecx^\herm(\bE-\bD)\vecx.
\end{align}
By H\"older's inequality, the second term satisfies
\begin{align} \label{eq:dictboundstep3}
\big|\vecx^\herm(\bE-\bD)\vecx\big| \leq \normone{\vecx}\norminf{(\bE-\bD)\vecx} \leq \coha\normone{\vecx}^2,
\end{align}
where the last step follows from
\begin{align}
\norminf{(\bE-\bD)\vecx} = \max_{i} \Big|-\coha x_i + \!\!\sum_{j\neq i}\!\veca_i^\herm\veca_jx_j\Big| \leq \max_i \Big\{\coha|x_i| +  \!\sum_{j\neq i}\!\coha |x_j|\Big\}  = \coha \normone{\vecx}.
\end{align}
Using \fref{eq:dictboundstep3} in \fref{eq:dictboundstep2}, excluding the case $\normtwo{\vecx}=0$, and dividing by $\normtwo{\vecx}^2$ yields
\begin{align}
\frac{\normtwo{\dicta\vecx}^2}{\normtwo{\vecx}^2} \geq 1+\coha - \coha\frac{\normone{\vecx}^2}{\normtwo{\vecx}^2} = 1-\coha(\sigma(\vecx)-1),
\end{align}
which is the lower bound in \fref{eq:dictbound2}. The upper bound follows analogously from 
\begin{align}
\normtwo{\dicta\vecx}^2 = \normtwo{\vecx}^2(1-\coha) + \vecx^\herm(\bE+\bD)\vecx \leq \normtwo{\vecx}^2(1-\coha) + \coha\normone{\vecx}^2,
\end{align} 
using the same bound as in \fref{eq:dictboundstep3} with the appropriate sign change.

\subsection[Proof of the Sigma-Density Kernel Result]{Proof of \fref{cor:sigmakernel}}
\label{app:sigmakernel}

If $\sigma(\vecx)<1+1/\coha$, then $1-\coha(\sigma(\vecx)-1)>0$, and the lower bound in \fref{eq:dictbound2} gives $\normtwo{\dicta\vecx}^2\geq\normtwo{\vecx}^2(1-\coha(\sigma(\vecx)-1))>0$.

\subsection[Proof of the Uncertainty Relation]{Proof of \fref{thm:uncertainty}}
\label{app:uncertainty}

To prove \fref{thm:uncertainty}, we will use the following result:
\begin{align} \label{eq:simplemutualcoherencebound}
\norminf{\dicta^\herm\dictb\vecz}& =\max_i\big|\veca^\herm_i\dictb\vecz\big|=\max_i\left|\sum_k\veca^\herm_i\vecb_k z_k\right|
\leq \max_i\sum_k\cohm|z_k| = \cohm \normone{\vecz}.
\end{align}
Analogously, we will use $\norminf{\dictb^\herm\dicta\vecx}\leq\cohm\normone{\vecx}$. Now, if $\dicta\vecx=\dictb\vecz$, then we have $\dicta^\herm\dicta\vecx=\dicta^\herm\dictb\vecz$ and also $\norminf{\dicta^\herm\dicta\vecx}=\norminf{\dicta^\herm\dictb\vecz}$. Using the LHS of  \fref{eq:normbound1} and \fref{eq:simplemutualcoherencebound}, we can  bound $\norminf{\dicta^\herm\dicta\vecx}$ from below and  $\norminf{\dicta^\herm\dictb\vecz}$ from above, respectively, and obtain
\begin{align} \label{eq:onebound1}
[1-\coha(\delta(\vecx)-1)]^+ \norminf{\vecx}\leq \cohm \normone{\vecz},
\end{align}
where we may take the non-negative part $[\cdot]^+$ on the LHS because the right-hand side (RHS) is non-negative, so that \fref{eq:onebound1} holds trivially whenever the lower bound from \fref{eq:normbound1} is negative.
Since $\dicta\vecx=\dictb\vecz$, we also have $\norminf{\dictb^\herm\dicta\vecx}=\norminf{\dictb^\herm\dictb\vecz}$. Using similar steps as above, we obtain 
\begin{align} \label{eq:onebound2}
[1-\cohb(\delta(\vecz)-1)]^+ \norminf{\vecz}\leq \cohm \normone{\vecx}.
\end{align}
Multiplying \fref{eq:onebound1} with \fref{eq:onebound2} and dividing both sides by $\norminf{\vecx}\norminf{\vecz}$  yields the uncertainty relation \fref{eq:uncertaintyprinciple}.

\section{Construction of Kernel Vectors with Low \texorpdfstring{$\gamma$}{gamma}-Density}
\label{app:gammaconstruction}

We detail the construction used in \fref{sec:whydelta}, which establishes that dictionaries with prescribed coherence possess kernel vectors whose $\gamma$-density is arbitrarily close to one.
Fix $0<\coha\leq1$, let $q\geq\max\{2,1/\coha\}$ be an integer, and set $M=q$ and $N_a=q+1$. Let $\vece_1\in\reals^{q}$ be the first standard unit vector, and let $\vecu_1,\ldots,\vecu_q\in\reals^{q}$ be unit vectors that are orthogonal to $\vece_1$ and form a regular simplex, i.e., $\sum_{j=1}^{q}\vecu_j=\bZero_{q\times1}$ and $\vecu_j^\herm\vecu_k=-1/(q-1)$ for $j\neq k$. Define the dictionary $\dicta\in\reals^{q\times(q+1)}$ with columns
\begin{align} \label{eq:gammaconstruction}
\veca_1=\vece_1 \quad \textnormal{and} \quad \veca_{j+1}=\coha\vece_1+\sqrt{1-\coha^2}\,\vecu_j, \,\, j=1,\ldots,q.
\end{align}
All columns have unit $\elltwo$-norm, $\veca_1^\herm\veca_{j+1}=\coha$, and $\veca_{j+1}^\herm\veca_{k+1}=(q\coha^2-1)/(q-1)$ for $j\neq k$, whose magnitude does not exceed $\coha$ whenever $q\coha\geq1$; hence, the coherence of $\dicta$ is~$\coha$. Furthermore, the spike-plus-uniform-tail vector
\begin{align} \label{eq:gammakernelvector}
\vecx=\Big[1,\,-\tfrac{1}{q\coha},\ldots,-\tfrac{1}{q\coha}\Big]^\herm
\end{align}
satisfies $\dicta\vecx=\bZero_{M\times1}$ with $\delta(\vecx)=1+1/\coha$ (meeting \fref{eq:kernelproofstep2} with equality) and $\gamma(\vecx)=1+1/(q\coha^2)\to1$ as $q=N_a-1\to\infty$. 
Since the kernel vector in \fref{eq:gammakernelvector} has a uniform tail and satisfies $\delta(\vecx)=1+1/\coha$ exactly, it furthermore attains both inequalities in \fref{eq:gammakernelbound} with equality.

\section{Proofs of \texorpdfstring{\fref{thm:OMPrecoveryA} and \fref{thm:OMPrecoveryB}}{the OMP Recovery Guarantees}}
\label{app:OMPproof}

\subsection{Preliminaries}

Both proofs share the following setup. Fix an iteration step, let $\setS$ denote the atoms selected by OMP so far, and partition the remaining indices $\setR=\{1,\ldots,N_a\}\backslash\setS$ into a set $\setG$ of admissible (``good'') and a set $\setB=\setR\backslash\setG$ of inadmissible (``bad'') candidates. OMP selects an admissible atom in this iteration whenever the following holds~\cite{donoho2006a}:
\begin{align} \label{eq:genericselection}
\max_{i\in \setG} \, \abs{\veca^\herm_i\bR_\setS \vecy} >  \max_{k\in\setB} \, \abs{\veca^\herm_k\bR_\setS \vecy}.
\end{align}
Here, $\bR_\setS=\bI_{M}-\bA_\setS\bA^\dagger_\setS$ is the projector onto the orthogonal complement of the atoms from the set~$\setS$. 
If $\setB=\varnothing$, then every remaining candidate is admissible, the iteration trivially selects an admissible atom, and there is nothing to prove; we therefore assume $\setB\neq\varnothing$ in what follows. The two proofs differ only in the choice of $\setG$ and $\setB$ and in the upper bound to be designed for the RHS of \fref{eq:genericselection}:
\begin{itemize}
\item For \fref{thm:OMPrecoveryB}, we sort the coefficients of $\vecx$ in descending order of magnitude (ties are resolved arbitrarily) and set $\setG=\setM_\setS$ and $\setB=\setR\backslash\setM_\setS$, where $\setM_\setS = \{ k\in\setR : |x_k|=\textstyle\max_{i\in\setR}|x_i| \}$ is the set of remaining indices of largest magnitude.
\item For \fref{thm:OMPrecoveryA}, we set $\setG=\setX\backslash\setS$ (the not-yet-recovered support) and $\setB=\setX^c$ (the off-support atoms), and proceed by induction with the hypothesis $\setS\subseteq\setX$.
\end{itemize}
For every remaining index $i\in\setR$, we can decompose $\vecy=\veca_ix_i + \bA_\setS\vecx_\setS+\bA_{\setR\backslash i}\vecx_{\setR\backslash i}$. Inserting this (index-dependent) decomposition into \fref{eq:genericselection} for each $i\in\setG$ on the LHS and for an arbitrary $i\in\setG$ on the RHS shows that \fref{eq:genericselection} is equivalent to
\begin{align} \label{eq:OMPcondition}
\max_{i\in \setG} \, \abs{\veca^\herm_i\bR_\setS(\veca_ix_i + \bA_\setS\vecx_\setS+\bA_{\setR\backslash i}\vecx_{\setR\backslash i})} >    \max_{k\in\setB} \, \abs{\veca^\herm_k\bR_\setS(\veca_ix_i + \bA_\setS\vecx_\setS+\bA_{\setR\backslash i}\vecx_{\setR\backslash i})}.
\end{align}
The fact that $\bR_\setS\bA_\setS=\bZero_{M\times\abs{\setS}}$ allows us to simplify \fref{eq:OMPcondition} to
\begin{align} \label{eq:OMPconditionsimpler}
&\max_{i\in \setG} \, \abs{\veca^\herm_i\bR_\setS(\veca_ix_i + \bA_{\setR\backslash i}\vecx_{\setR\backslash i})} >  \max_{k\in\setB} \, \abs{\veca^\herm_k\bR_\setS\bA_\setR\vecx_\setR}.
\end{align}
To arrive at a sufficient condition that guarantees success, we now individually lower- and upper-bound the LHS and RHS of \fref{eq:OMPconditionsimpler}, respectively. The lower bound on the LHS is common to both proofs; the upper bounds on the RHS differ and are derived in \fref{app:OMPproofB} and \fref{app:OMPproofA}.

\subsection[Lower Bound on the LHS of the OMP Selection Condition]{Lower Bound on the LHS of \fref{eq:OMPconditionsimpler}}

Let $\hat i\in\arg\max_{i\in\setG}\abs{x_i}$ denote (one of) the admissible indices with largest coefficient magnitude, fixed for the remainder of this derivation. 
For $\setS=\varnothing$ (i.e., in the first OMP iteration), we have $\bR_\setS=\bI_M$, all terms involving $\bA_\setS$ below vanish, and the resulting bounds hold trivially; in what follows, we therefore assume $\setS\neq\varnothing$, so that the inverse in \fref{eq:weirdmethod} is well-defined. Since the maximum over $\setG$ dominates the term associated with $\hat i$, and since $\bR_\setS\bA_\setS=\bZero_{M\times\abs{\setS}}$, we obtain
\begin{align} \label{eq:LHSexpansion}
\max_{i\in \setG} \, \abs{\veca^\herm_i\bR_\setS\vecy} \geq \abs{\veca^\herm_{\hat i}\bR_\setS\vecy} = \abs{x_{\hat i} + \veca^\herm_{\hat i}\bA_{\setR\backslash \hat i}\vecx_{\setR\backslash \hat i}  -\veca^\herm_{\hat i}\bA_\setS\bA^\dagger_\setS(\veca_{\hat i}x_{\hat i} + \bA_{\setR\backslash \hat i}\vecx_{\setR\backslash \hat i})}.
\end{align}
Applying the reverse and regular triangle inequalities, together with $\abs{\veca^\herm_{\hat i}\veca_k}\leq\coha$ for $k\neq\hat i$, yields
\begin{align} \label{eq:LHSlowerbound}
\abs{\veca^\herm_{\hat i}\bR_\setS\vecy} \, & \geq \abs{x_{\hat i}} -\abs{\veca^\herm_{\hat i}\bA_{\setR\backslash \hat i}\vecx_{\setR\backslash \hat i}}  -\abs{\veca^\herm_{\hat i}\bA_\setS\bA^\dagger_\setS(\veca_{\hat i}x_{\hat i} + \bA_{\setR\backslash \hat i}\vecx_{\setR\backslash \hat i})} \notag \\
& \geq \abs{x_{\hat i}}(1+\coha) - \sum_{k\in \setR} \coha \abs{x_k} - \sum_{k \in \setR} \abs{\veca^\herm_{\hat i} \bA_\setS\bA^\dagger_\setS\veca_k}\abs{x_k }.
\end{align}
We now bound the last term in the above result  as follows:
\begin{align}
 \abs{\veca^\herm_{\hat i} \bA_\setS\bA^\dagger_\setS\veca_k}   \leq \| \bA_\setS^\herm\veca_{\hat i}\|_2\|(\bA^\herm_\setS\bA_\setS)^{-1}\bA^\herm_\setS\veca_k\|_2  
&  \leq  \frac{\normtwo{\bA_\setS^\herm \veca_{\hat i} }\normtwo{\bA^\herm_\setS\veca_k}}{[1-\coha(\abs{\setS}-1)]^+}  \leq  \frac{\coha^2\abs{\setS}}{[1-\coha(\abs{\setS}-1)]^+}, \label{eq:weirdmethod}
\end{align}
where we used the Cauchy--Schwarz inequality, Ger\v{s}gorin's disc theorem \cite[Thm.~6.1.1]{hornjohnson}, and the definition of the dictionary coherence $\coha$ in \fref{eq:coherence}, exploiting that $\hat i,k\in\setR$ are distinct from all indices in $\setS$.
We note that \fref{eq:weirdmethod} requires $\abs{\setS} < 1/\coha+1$, which must hold for any set size picked by OMP; as a result, we get condition \fref{eq:OMPfirstcondition}.
By combining the above bounds and using $\abs{x_{\hat i}}=\max_{i\in\setG}\abs{x_i}$, we arrive at
\begin{align}
&\max_{i\in \setG} \, \abs{\veca^\herm_i\bR_\setS\vecy} \geq  \max_{i\in \setG} \, \abs{x_i}(1+\coha) - \frac{\coha+\coha^2}{[1-\coha(\abs{\setS}-1)]^+}\sum_{k\in \setR} \abs{x_k}.
\label{eq:OMPlowerboundcondition}
\end{align}

\subsection[Proof of the Largest-Entries Identification Theorem]{Proof of \fref{thm:OMPrecoveryB}}
\label{app:OMPproofB}

Here, $\setG=\setM_\setS$ is the set of remaining indices of largest magnitude, as defined in the preliminaries above, and $\setB=\setR\backslash\setM_\setS$, so every $k\in\setB$ satisfies $\abs{x_k}\leq m_2$, where $m_2$ is the largest remaining magnitude strictly below $m_1=\max_{i\in\setR}\abs{x_i}$ (i.e., $m_1$ and $m_2$ correspond to $m_1^{(t)}$ and $m_2^{(t)}$ in \fref{thm:OMPrecoveryB}, with the iteration index omitted for brevity). 
Note that $t\leq t_\textnormal{max}\leq\normzero{\vecx}$ together with the induction hypothesis (only non-zero entries have been selected so far) ensures that at least one non-zero entry remains, so $m_1>0$. We upper-bound the RHS of \fref{eq:OMPconditionsimpler}. Expanding and applying the triangle inequality gives
\begin{align}
\abs{\veca^\herm_k\bR_\setS\bA_\setR\vecx_\setR} \leq \abs{\veca^\herm_k\bA_\setR\vecx_\setR} + \abs{\veca^\herm_k\bA_\setS\bA^\dagger_\setS\bA_\setR\vecx_\setR}.
\end{align}
Because $k\in\setR$, the first term retains the diagonal contribution $\veca^\herm_k\veca_kx_k=x_k$; isolating it results in
\begin{align}
\abs{\veca^\herm_k\bA_\setR\vecx_\setR} = \Big|x_k + \!\!\sum_{j\in\setR\backslash k}\!\! \veca^\herm_k\veca_jx_j\Big| \leq \abs{x_k} + \coha\!\!\sum_{j\in\setR\backslash k}\!\!\abs{x_j} \leq m_2 + \coha\!\sum_{j\in\setR}\!\abs{x_j}.
\end{align}
The second term is bounded as in \fref{eq:weirdmethod}. Combining the two and using 
\begin{align}
\coha+\coha^2\abs{\setS}/[1-\coha(\abs{\setS}-1)]^+ = (\coha+\coha^2)/[1-\coha(\abs{\setS}-1)]^+, 
\end{align}
we obtain
\begin{align} \label{eq:OMPupperboundB}
\max_{k\in\setB} \, \abs{\veca^\herm_k\bR_\setS \vecy} \leq m_2 + \frac{\coha+\coha^2}{[1-\coha(\abs{\setS}-1)]^+}\!\sum_{j\in \setR} \abs{x_j}.
\end{align}
Together with the LHS bound in \fref{eq:OMPlowerboundcondition}, in which $\max_{i\in\setG}\abs{x_i}=m_1$, a sufficient condition for \fref{eq:OMPconditionsimpler} is
\begin{align}
m_1(1+\coha) - m_2 > \frac{2\coha(1+\coha)}{[1-\coha(\abs{\setS}-1)]^+}\sum_{k\in \setR} \abs{x_k}.
\end{align}
Dividing by $m_1(1+\coha)$ and using $\delta(\vecx_\setR)=\sum_{k\in\setR}\abs{x_k}/m_1$, $\abs{\setS}=t-1$, and $\rho^{(t)}=m_2/m_1$ gives
\begin{align}
\delta(\vecx_\setR) < \tfrac{1}{2}\big({1}/{\coha}+1-(t-1)\big)\Big(1-\tfrac{\rho^{(t)}}{1+\coha}\Big),
\end{align}
which is \fref{eq:OMPconditionB}. If \fref{eq:OMPfirstcondition} is met, we have $[1-\coha(\abs{\setS}-1)]^+>0$ throughout, so the  pseudo-inverse is guaranteed to exist. By induction over $t$, the support set $\setS^{(t)}$ collects the indices of the $t$ largest entries of~$\vecx$; if $\normzero{\vecx}=t_\textnormal{max}$, OMP recovers $\vecx$ exactly.

\subsection[Proof of the Strictly-Sparse Recovery Theorem]{Proof of \fref{thm:OMPrecoveryA}}
\label{app:OMPproofA}

We argue by induction and assume $\setS=\setS^{(t-1)}\subseteq\setX$ with $\abs{\setS}=t-1$; the base case $\setS^{(0)}=\varnothing$ is trivial. With $\setG=\setX\backslash\setS$ and $\setB=\setX^c$, every inadmissible index $k\in\setB$ lies outside the support. Since $x_j=0$ for $j\in\setR\backslash\setG$, we have $\bA_\setR\vecx_\setR=\bA_\setG\vecx_\setG$, and because $k\notin\setG$ the first RHS term carries no diagonal contribution, which leads to
\begin{align}
\abs{\veca^\herm_k\bA_\setR\vecx_\setR} = \Big|\sum_{j\in\setG} \veca^\herm_k\veca_jx_j\Big| \leq \coha\sum_{j\in\setG}\abs{x_j}.
\end{align}
Bounding the second term as in \fref{eq:weirdmethod} and combining yields
\begin{align} \label{eq:OMPupperboundA}
\max_{k\in\setB} \, \abs{\veca^\herm_k\bR_\setS \vecy} \leq \frac{\coha+\coha^2}{[1-\coha(\abs{\setS}-1)]^+}\!\sum_{j\in \setG} \abs{x_j}.
\end{align}
Combining \fref{eq:OMPupperboundA} with \fref{eq:OMPlowerboundcondition} (where $\sum_{k\in\setR}\abs{x_k}=\sum_{j\in\setG}\abs{x_j}$ and $\max_{i\in\setG}\abs{x_i}=m_1$), a sufficient condition for~\fref{eq:OMPconditionsimpler} is therefore
\begin{align}
m_1(1+\coha) > \frac{2\coha(1+\coha)}{[1-\coha(\abs{\setS}-1)]^+}\sum_{j\in \setG} \abs{x_j}.
\end{align}
By dividing by $m_1(1+\coha)$ and using $\delta(\vecx_\setG)=\sum_{j\in\setG}\abs{x_j}/m_1$ with $\setG=\setX\backslash\setS^{(t-1)}$ and $\abs{\setS}=t-1$, we obtain the condition
\begin{align}
\delta\big(\vecx_{\setX\backslash\setS^{(t-1)}}\big) < \tfrac{1}{2}\big({1}/{\coha}+1-(t-1)\big),
\end{align}
which corresponds to~\fref{eq:OMPconditionA}. Hence OMP selects an atom from $\setX$, i.e., $\setS^{(t)}\subseteq\setX$, which completes the induction. After $s$ iterations, $\setS^{(s)}=\setX$ and OMP recovers $\vecx$ exactly.


\bibliographystyle{IEEEtran} 

\bibliography{IEEEabrv,confs-jrnls,publishers,studer}

\end{document}